\documentclass[a4paper,11pt]{article}
\pdfoutput=1 

\usepackage{jcappub} 

\usepackage[T1]{fontenc} 
\usepackage{placeins}
\usepackage[dvipsnames,svgnames,x11names,hyperref]{xcolor}
\usepackage{array}
\usepackage{graphicx}
\usepackage{booktabs}
\usepackage{slashed} 
\RequirePackage{mathrsfs}
\usepackage{dsfont}
\usepackage{hyperref}
\hypersetup{
        unicode = true,
        pdftitle = Note , 
        colorlinks = true,
        linkcolor = Black!30!DodgerBlue4,
        citecolor = Black!30!DodgerBlue4,
        filecolor = Black!30!DodgerBlue4,
        urlcolor = Black!30!DodgerBlue4,
}

\usepackage{booktabs}
\usepackage{cancel} 
\graphicspath{{./plots/}}

\usepackage{floatrow}
\newfloatcommand{capbtabbox}{table}[][\FBwidth]

\usepackage{ragged2e}

\renewcommand{\epsilon}{\varepsilon}

\renewcommand{\vec}[1]{\boldsymbol{#1}}
\renewcommand{\bar}{\overline}

\renewcommand{\Im}{\mathrm{Im}}

\newcolumntype{L}{>{$}l<{$}}
\newcolumntype{C}{>{$}c<{$}}

\allowdisplaybreaks

\title{\boldmath \justifying On the dark matter origin of an LDMX signal}

\author[a]{Riccardo Catena}
\author[a]{Taylor R. Gray}
\author[a]{and Andreas Lund}

\affiliation[a]{Chalmers University of Technology, Department of Physics, SE-412 96 G\"oteborg, Sweden}

\emailAdd{catena@chalmers.se}
\emailAdd{taylor.gray@chalmers.se}
\emailAdd{lundand@chalmers.se}

\abstract{Fixed target experiments where beam electrons are focused upon a thin target have shown great potential for probing new physics, including the sub-GeV dark matter (DM) paradigm. However, a signal in future experiments such as the light dark matter experiment (LDMX) would require an independent validation to assert its DM origin. To this end, we propose to combine LDMX and next generation DM direct detection (DD) data in a four-step analysis strategy, which we here illustrate with Monte Carlo simulations. In the first step, the hypothetical LDMX signal (i.e.~an excess in the final state electron energy and transverse momentum distributions) is {\it recorded}. In the second step, a DM DD experiment operates with increasing exposure to test the DM origin of the LDMX signal. Here, LDMX and DD data are simulated. In the third step, a posterior probability density function (pdf) for the DM model parameters is extracted from the DD data, and used to {\it predict} the electron recoil energy and transverse momentum distributions at LDMX. In the last step, {\it predicted} and {\it recorded} electron recoil energy and transverse momentum distributions are compared in a chi-square test. We present the results of this comparison in terms of a threshold exposure that a DD experiment has to operate with to assert whether {\it predicted} and {\it recorded} distributions {\it can} be statistically dependent. We find that this threshold exposure grows with the DM particle mass, $m_\chi$. It varies from 0.0013~kg-year for a DM mass of $m_\chi=4$~MeV to 1.9~kg-year for $m_\chi=25$ MeV, which is or will soon be within reach.} 

\begin{document} 
\maketitle
\flushbottom

\section{Introduction}

There is mounting interest in dark matter (DM) models where the DM particle is lighter than a GeV, and lies in the same mass range of the known constituents of matter, such as electrons, protons and neutrons~\cite{Battaglieri:2017aum,Mitridate:2022tnv,Balan:2024cmq}.~The reason for this increased interest is twofold. On the one hand, a DM candidate in this mass range would carry a kinetic energy smaller than $\sim 10^{-6} m_\chi$ in our galaxy, where $m_\chi$ is the DM mass, and would thus ``by construction'' evade the increasingly strong constraints from direct DM searches in nuclear recoil experiments~\cite{Billard:2021uyg}. On the other hand, a DM candidate in this mass range could also be produced in the early universe via the canonical freeze-out mechanism~\cite{Fabbrichesi:2020wbt}, as long as the relevant number-changing processes involve the exchange of a new particle mediator, so that the Lee-Weinberg bound~\cite{Lee:1977ua} (that applies to weak-scale processes) can be circumvented. This latter observation in particular implies that the search for sub-GeV DM is tightly related to the search for new mediator particles.

New particle mediators can be searched for at fixed target experiments, where a beam of particles (electrons or protons) collides with a target at rest, producing the new mediator in, e.g.: 1) ``dark'' bremsstrahlung processes, where the new particle replaces the ordinary final state photon, or 2) meson decays, where the new mediator is produced in $\pi^0$ or $\eta$ decays~\cite{deNiverville:2016rqh}. Once produced, the new mediator particle can decay visibly into Standard Model particles, or invisibly into DM particles, which illustrates the aforementioned interplay between mediator and DM particle searches in the sub-GeV DM context. In the next generation fixed target experiment LDMX~\cite{LDMX:2018cma}, 4 -- 8 GeV beam electrons are focused upon a thin tungsten target, and the hypothetically produced new particle mediator is searched for by full reconstruction of the final state electron recoil energy and transverse momentum distributions. The prospects for new mediator particle or DM production at LDMX are reviewed in \cite{Akesson:2022vza} (and references therein), and a recent extension of the LDMX analysis framework to spin-1 DM is discussed in \cite{Catena:2023use}.

Importantly, if an excess above the expected inclusive single electron background is found at LDMX in the recorded electron recoil energy and transverse momentum distributions, there is no guarantee that this excess is related to DM. Indeed, from this LDMX signal one could only infer that an invisible particle carrying a significant fraction of the initial beam energy has been produced. Whether or not this new particle is associated with DM, and plays a key role in the DM cosmological production is a question that would have to be addressed separately.
Complementarity between accelerator and direct detection experiments for Weakly Interacting Massive Particles (WIMPs) has been explored previously in \cite{Baum:2018lua}.

In this work, we propose a strategy that can be used to test the DM origin of a future LDMX signal. Besides the LDMX signal itself, our proposal relies on information that can be extracted from next generation DM direct detection experiments, and consists of four steps. In the first step, the hypothetical LDMX signal is recorded (or, to illustrate our approach, simulated). In the second step a DM direct detection experiment operates with increasing exposure, and parameter inference is performed on the recorded data to extract posterior probability density function (pdf) and associated credible regions for the DM mass and coupling constants.~Here, we perform Monte Carlo simulations of electron/hole pair production in a silicon target to emulate the operation of a next generation, DM direct detection experiment. In the third step, the posterior pdf obtained from DM direct detection data is used to predict the electron recoil energy and transverse momentum distributions that are expected at LDMX based on the outcome of the DM direct detection experiment in question. Obviously, the predicted distributions will depend on the assumed experimental exposure. In the last step, predicted and observed electron recoil energy and transverse momentum distributions are compared in a chi-square test to determine whether the two distributions have been sampled from different underlying models. This last step allows us to find the exposure that is required to assert the compatibility of the hypothetical LDMX signal with a DM origin. This strategy highlights the strong complementarity of mediator particle searches at fixed target experiments and DM searches at direct detection experiments.
An alternative approach consists of using the transverse momentum and electron recoil energy distributions measured at LDMX to predict the electron-hole spectrum expected at future DM direct detection experiments. One advantage of our strategy compared to the latter is that it enables us to compare the ``observed'' LDMX transverse momentum and electron recoil energy distributions with the two analogous distributions extracted from direct detection data. Had we followed the complementary (and computationally simpler) approach of comparing an LDMX prediction with direct detection observations, we would have been able to compare only one type of distribution: the electron-hole distributions predicted by LDMX and observed at direct detection.

Our work is organised as follows. In Sec.~\ref{sec:theory} we review the particle physics model used to describe DM in our analysis strategy (the four steps of which were outlined above). In Sec.~\ref{sec:sim} and Sec.~\ref{sec:stat} we introduce the simulation and statistical frameworks used to implement our strategy. The numerical results are presented in Sec.~\ref{sec:results}, while we summarise and conclude in Sec.~\ref{sec:conclusions}.

\section{Theoretical framework}
\label{sec:theory}

In order to illustrate our proposal, we extend the Standard Model Lagrangian, $\mathscr{L}_{\rm SM}$, as follows~\cite{Holdom:1985ag,Babu:1997st,Berlin:2018bsc}
\begin{align}
\mathscr{L} = \mathscr{L}_{\rm SM} -\frac{1}{4} F'_{\mu \nu} F'^{\mu\nu}-\frac{1}{2} m_{A'} A'_{\mu} A'^{\mu} - A'_\mu \left( \epsilon e J^\mu_{\rm SM} + g_{\rm D}J^\mu_{\rm D}  \right) \,,
\label{eq:L}
\end{align}
where $A'_\mu$ is the gauge boson of a new $U(1)$ gauge group with field strength tensor $F'_{\mu \nu}=\partial_\mu A'_\nu - \partial_\nu A'_\mu$ and mass $m_{A'}$. The Lagrangian in Eq.~(\ref{eq:L}) is valid at energies below the electroweak symmetry breaking scale, and assumes that the vector field $A'_\mu$ has acquired a mass either through the Higgs mechanism (which implies potentially relevant couplings of $A'_\mu$ with an additional Higgs field not included in Eq.~(\ref{eq:L})) or via the Stueckelberg mechanism (where $A'_\mu$ is manifestly decoupled from other scalar degrees of freedom) \cite{Fabbrichesi:2020wbt}. The Lagrangian in Eq.~(\ref{eq:L}) also assumes that the fields have been transformed to mass eigenstates, and $m_{A'}\ll m_{Z}$, where $m_{Z}$ is the $Z$-boson mass. The vector field $A'_\mu$ introduces new particles referred to as ``dark photons''.

The coupling between the dark photon and the charged fermions of the Standard Model arises from a kinetic mixing between $A'_\mu$ and the hypercharge field, and is associated with the current,
\begin{align}
J^\mu_{\rm SM} = \sum_{f} Q_f\, \bar{f} \gamma^\mu f \,,
\end{align}
where the sum runs over the Standard Model fermions of electric charge $Q_f$ and Dirac spinors $f$. In Eq.~(\ref{eq:L}), the kinetic mixing parameter is denoted by $\epsilon$, while $e$ is the elementary charge. 
The coupling between the dark photon field and the DM field, $\chi$, is associated with the current
\begin{align}
J^{\mu}_{\rm D} = i \left( \chi^* \partial^\mu \chi - \chi \partial^\mu \chi^* \right) \,,
\end{align}
where we assumed that $\chi$ is a complex scalar. The corresponding coupling constant is denoted by $g_{\rm D}$.\\

In the case of complex scalar DM, angular momentum conservation implies that the thermally averaged, s-channel annihilation cross section is $p$-wave, and thus suppressed by two powers of the DM speed in the non-relativistic limit. Consequently, DM can be thermally produced in the early universe, while constraints from exotic energy injection into the photon-baryon plasma prior recombination and from X-ray measurements are basically absent~\cite{Boehm:2003hm,Boehm:2020wbt}. At the same time, Eq.~(\ref{eq:L}) implies a velocity-independent coupling between DM and the nucleon and electron densities in materials, which can therefore be probed with DM direct detection experiments \cite{Catena:2022fnk}.\\

We refer to Ref.~\cite{Balan:2024cmq} for a comprehensive global analysis of complex scalar DM within {\sffamily GAMBIT}.

\section{Simulation framework}
\label{sec:sim}

In this section, we introduce the simulation framework we apply to test the DM origin of an LDMX signal. Specifically, Sec.~\ref{sec:ldmx} introduces the framework we use to simulate a hypothetical LDMX signal from a point in parameter space where DM has the correct relic density. Sec.~\ref{sec:dd} focuses on the tools we employ to simulate a future DM direct detection experiment operating with increasing exposure. Finally, Sec.~\ref{sec:ldmx2} revisits our LDMX computations, and describes how we translate the recorded data of a future DM direct detection experiment into a prediction for the electron recoil energy and transverse momentum distributions at LDMX.

\subsection{LDMX: hypothetical signal}
\label{sec:ldmx}

In order to simulate a signal at LDMX, i.e. an excess in the electron recoil energy and transverse momentum distributions over the expected inclusive single electron backgrounds, we employ the Monte Carlo event generator {\sffamily MadGraph5\_aMC@NLO}~\cite{Alwall:2011uj}. Specifically, we use {\sffamily MadGraph} to simulate the process $e^-W\rightarrow e^- W A'\rightarrow e^- W \chi \chi$, where a dark photon is produced in electron-tungsten collisions, and then invisibly decays into a pair of DM particles. In experiments exploiting an electron beam, such as LDMX, vector meson decays significantly contribute to the experimental sensitivity only for DM particle masses above 100~MeV~\cite{Schuster:2021mlr}, and we thus neglect this contribution to dark photon production here.~In the simulations, we set the beam energy to 4 GeV, and model tungsten as an ``elementary'' particle of spin-1/2, mass 171 GeV, and electric charge +74, coupling to the ordinary photon via the vertex $i G(t) \gamma^\mu$, where~\cite{Bjorken:2009mm} 
\begin{align}
G(t) &= G_{2, \textrm{el}}(t) + G_{2, \textrm{in}}(t) \,,
\end{align}
is a nuclear form factor receiving an elastic contribution (which dominates for most masses considered here),
\begin{align}
G_{2, \textrm{el}}(t)= \left( \frac{a^2t}{1+a^2t} \right)^2 \left(  \frac{1}{1+t/d} \right)^2 Z^2
\end{align}
and an inelastic contribution,
\begin{align}
G_{2, \textrm{in}}(t)= \left( \frac{a^{\prime2}t}{1+a^{\prime2}t} \right)^2 \left[  \frac{1+\frac{t}{4m_p^2}\left( \mu_p^2-1 \right)}{\left(1+\frac{t}{0.71~\textrm{GeV}}\right)} \right] Z\,.
\end{align}
Here, $\gamma^\mu$ denotes a $4\times4$ gamma-matrix, $d=0.164$~GeV$^2$~$A^{-2/3}$, $A=184$, $a=111~Z^{-1/3}/m_e$, $a^\prime=773~Z^{-2/3}/m_e$, $Z=74$, $\mu_p=2.79$~GeV, $m_p$ is the proton mass, and $t=-(P_i-P_f)^2$, $P_i$ and $P_f$ being the initial and final tungsten four momenta, respectively. Furthermore, we assume a thickness (density) of 0.035 cm (19.3 g~cm$^{-3}$) for the tungsten target, an integrated luminosity $L$ corresponding to $4\times10^{14}$ electrons on target (EOT)\footnote{Specifically, $L$ is given by $0.1 \times X_0 \times \textrm{EOT}\times N_0 / A$ where $0.1 X_0$ stands for 10\% radiation length, and $N_0/A$ is the number of particles per gram of material. See~\cite{LDMX:2018cma} for further details.}, and a 50\% experimental efficiency, $\eta=0.5$ 

The above information is encoded into a UFO file, which extends the Standard Model by a ``tungsten particle'' (modelled as explained above), a dark photon, and a scalar DM particle with interactions that can be extracted from the Lagrangian in Eq.~(\ref{eq:L}). Following~\cite{LDMX:2018cma}, we also assume the relation $m_{A'}=3 m_\chi$, which implies that, once produced, the dark photon decays invisibly into a pair of DM particles with branching ratio close to 1. Finally, we set $\alpha_{\rm D}=0.5$.

In all simulations, we treat $m_\chi$ and $\epsilon$ as free parameters. The input number of $e^-W\rightarrow e^- W A'\rightarrow e^- W \chi \chi$ events provided by the user to {\sffamily MadGraph}, $N_{\rm MG}$, is determined by the choice of $m_\chi$ and $\epsilon$, namely
\begin{align}
N_{\rm MG}(m_\chi, \epsilon)=\eta L \sigma_{\rm ee}(m_\chi, \epsilon)\,, 
\end{align}
where $\sigma_{\rm ee}(m_\chi, \epsilon)$ is the cross section for the $e^-W\rightarrow e^- W A'\rightarrow e^- W \chi \chi$ process at $m_\chi$ and $\epsilon$. 

We distribute the $N_{\rm MG}$ events recorded in a {\sffamily MadGraph} run in $n_b=30$ bins for the final state electron recoil energy, $E_e$, and $n_b=30$ bins for the final state electron transverse momentum $P_T\equiv|\vec{P}_T|$, and denote the corresponding number of counts per bin by $N^{E_e}_{i,\textrm{MG}}$ and $N^{P_T}_{i,\textrm{MG}}$, $i=1,\dots n_b$, respectively. The bins are constructed up to a maximum energy or transverse momentum threshold, so that each bin has sufficient statistics (i.e. more than 5 samples). Notice that we use the symbol $N_{i,\textrm{MG}}$ for both distributions, and let the labels $E_e$ and $P_T$ specify the type of counts. Notice also that\footnote{This relation is true to good approximation, since only a few energies or transverse momenta are discarded when constructing histograms with at least 5 counts per bin.}
\begin{align}
&\sum_{i=1}^{n_b} N^{E_e}_{i,\textrm{MG}} = N_{\rm MG} \,, \nonumber\\
&\sum_{i=1}^{n_b} N^{P_T}_{i,\textrm{MG}} = N_{\rm MG} \,.
\end{align}
It is computationally cumbersome to run a new Madgraph simulation for many choices of $m_\chi$ and $\epsilon$, which is particularly detrimental when determining the sensitivity of the LDMX result to the uncertainties from a parameter estimation of $m_\chi$ and $\epsilon$. In order to circumvent this, we are interested in finding an analytical expression that relates $N^{E_e}_{i,\textrm{MG}}(m_\chi, \epsilon)$ and $N^{P_T}_{i,\textrm{MG}}(m_\chi, \epsilon)$ to a reference run with $N^{E_e}_{i,\textrm{MG}}(m_{\chi}, \epsilon_{\rm ref})$ and $N^{P_T}_{i,\textrm{MG}}(m_{\chi}, \epsilon_{\rm ref})$ counts. Here, $\epsilon_{\rm ref}$ is a reference value for the kinetic mixing parameter at which we evaluate $\sigma_{\rm ee}$ using {\sffamily MadGraph} for each given $m_\chi$. Focusing on the number of counts $N^{E_e}_{i}(m_\chi, \epsilon)$, we find the following relation, 
\begin{align}
N^{E_e}_{i,\textrm{MG}}(m_\chi, \epsilon) &= N^{E_e}_{i,\textrm{MG}}(m_\chi, \epsilon_{\rm ref}) \frac{N_{\textrm{MG}}(m_\chi, \epsilon)}{N_{\rm MG}(m_\chi, \epsilon_{\rm ref})} \nonumber\\
&=N^{E_e}_{i,\textrm{MG}}(m_\chi, \epsilon_{\rm ref}) \frac{\eta L \epsilon^2 \sigma_{\rm ee}(m_\chi,\epsilon_{\rm ref})}{\epsilon^2_\textrm{ref}N_{\rm MG}(m_\chi, \epsilon_{\rm ref})} \nonumber\\
&=
\frac{\epsilon^2}{\epsilon^2_{\rm ref}}\,N^{E_e}_{i,\textrm{MG}}(m_\chi, \epsilon_{\rm ref}) \,,
\label{eq:Ee}
\end{align}
where we used the fact that $\sigma_{\rm ee}(m_\chi,\epsilon)$ depends quadratically on $\epsilon$.
Notice that $N^{E_e}_{i,\textrm{MG}}(m_\chi, \epsilon)$ only depends on $\epsilon$ through the $\epsilon^2$ factor in Eq.~(\ref{eq:Ee}). A relation analogous to Eq.~(\ref{eq:Ee}) applies also to $N^{P_T}_{i,\textrm{MG}}(m_\chi, \epsilon)$.

Within this framework, we simulate a hypothetical LDMX signal by running {\sffamily MadGraph} with the UFO file described above, setting $m_\chi=m_\chi^*$ and $\epsilon=\epsilon^*$, where $m_\chi^*$ and $\epsilon^*$ are values of the DM mass and kinetic mixing parameter that correspond to the correct DM relic density. For each given $m_\chi^*$, we read the corresponding value of $\epsilon^*$ from Fig. 5 of \cite{LDMX:2018cma}. 

We denote the electron recoil energy and transverse momentum distributions representing the simulated LDMX signal by $N^{E_e}_{\textrm{LDMX},i}$ and $N^{P_T}_{\textrm{LDMX},i}$, $i=1,\dots, n_b$, respectively. Since {\sffamily MadGraph} is a Monte Carlo generator, there is an intrinsic error associated with our simulated signal, which is due to the finite size of our Monte Carlo samples. We estimate this error by simulating the distributions $N^{E_e}_{\textrm{LDMX},i}$ and $N^{P_T}_{\textrm{LDMX},i}$, $i=1,\dots, n_b$, 10 times for each given ($m_\chi^*$, $\epsilon^*$) pair. We denote the resulting Monte Carlo errors by $\sigma^{E_e}_{\rm MG}$ and $\sigma^{P_T}_{\rm MG}$, respectively.

In our analysis, we have distributed the LDMX simulated events in 30 recoil energy and transverse momentum bins. This choice is a trade off between resolution and computational cost. We find that a lower number of bins would make our computations slightly faster but the resolution of the recoil energy and transverse momentum distributions would be poor. On the other hand, a higher number of bins would lead to smoother, high-resolution histograms, at the expense of increased computational time. With this trade-off in mind, we determined that 30 bins gives
a satisfactory accuracy for our purposes.

\subsection{Direct detection: tests with increasing exposure}
\label{sec:dd}

Having generated a positive signal at LDMX from the benchmark point ($m_\chi^*$, $\epsilon^*$) as explained in Sec.~\ref{sec:ldmx}, we now proceed by simulating a next generation DM direct detection experiment that operates with increasing exposure to test the DM origin of the hypothetical LDMX signal. 

To this end, we employ the {\sffamily DarkELF} computer program~\cite{Knapen:2021bwg}. DarkELF calculates interaction rates of light DM in dielectric materials, including screening effects. Detector material properties are parametrised in terms of the dielectric function, $\epsilon_r$, which is pre-computed and tabulated for a number of target materials by using the density-functional theory code {\sffamily GPAW}~\cite{Mortensen_2024}, which implements the projector-augmented wave method. Here we focus on silicon as a target material, and restrict ourselves to the calculation of electronic transitions induced by DM-electron scattering in the detector. We therefore neglect DM absorption and the Migdal effect, which are found to be sub-leading in the DM mass range studied in this work.

In terms of the dielectric function, the rate per unit detector mass of electronic transitions induced by the scattering of DM particles by the electrons bound to a detector material takes the following form
\begin{align}
\frac{{\rm d}\mathscr{R}}{{\rm d} \omega} = \frac{1}{2 \pi \rho_T} \frac{\rho_\chi}{m_\chi} \frac{\sigma_e}{\mu_{\chi e}^2}\frac{\pi}{\alpha} \int {\rm d}^3\vec v \, f_\chi(\vec v) \int \frac{{\rm d}^3\vec q}{(2\pi)^3} \, q^2 |F_{\rm DM}(\vec q)|^2  \, \Im \left( -\frac{1}{\epsilon_r(\vec q,\omega)} \right)  \, \delta\left(\omega + \Delta E_\chi \right)\,,
\label{eq:dR/domega}
\end{align}
where $\rho_T$ is the target density, $m_\chi$ is the DM particle mass, $\rho_\chi$ is the local DM density, $f_\chi(\vec v)$ is the local DM velocity distribution in the laboratory frame, $\Delta E_\chi = q^2/(2 m_\chi) -\vec q \cdot \vec v$, $\mu_{\chi e}$ is the reduced DM-electron mass, $\vec q$ is the momentum transferred from the DM to the target, $\omega$ is the deposited energy, $\alpha$ is the fine structure constant, and $\sigma_e$ is a reference cross section given by
\begin{align}
\sigma_e = \frac{16 \pi  \alpha \alpha_{\rm D} \, \epsilon^2 \mu_{\chi e}^2 }{(m_{A'}^2 + \alpha^2 m_e^2)^2} \,,
\label{eq:sigmae}
\end{align}
where $\alpha_{\rm D}=g_{\rm D}^2/(4 \pi)$. In all numerical applications below, we will assume $\alpha_D=0.5$. Finally,
\begin{align}
F_{\rm DM}(\vec q) = \frac{m_{A'}^2 + \alpha^2 m_e^2}{q^2 + m_{A'}^2} \,,
\end{align}
is the so-called DM form factor. For $m_{A'}^2 \gg   \alpha^2 m_e^2$ and $m_{A'}^2 \gg q^2$, $F_{\rm DM}(\vec q) =1$, corresponding to the heavy mediator case. In the light/massless mediator case, instead, one has $F_{\rm DM}(\vec q) = \alpha^2 m_e^2/q^2$.  In the context of DM direct detection, we will use Eq.~(\ref{eq:sigmae}) and adopt $\sigma_e$, rather than $\epsilon$, as free parameter. The two approaches are equivalent, as long as one assumes $m_{A'}=3 m_\chi$. 

The dielectric function in Eq.~(\ref{eq:dR/domega}) is the longitudinal part of the dielectric tensor in units of the vacuum permittivity. In linear response theory, the dielectric tensor arises as the proportionality factor between an external electric field, and the induced displacement vector \cite{Catena:2024rym}.

In our simulations, we are interested in the number of electronic transitions, i.e. events, where an integer number $Q$ of electron/hole pairs is produced. This is obtained from Eq.~(\ref{eq:dR/domega}) as follows 
\begin{align}
 \frac{{\rm d}N(Q)}{{\rm d} Q} = \xi \int_{\Delta \omega(Q)} {\rm d}\omega \, \frac{{\rm d}\mathscr{R}(\omega)}{{\rm d} \omega} \,,
\end{align}
where $\xi$ is the experimental exposure (detector mass $\times$ data taking time), whereas $\Delta \omega(Q)$ is the range of deposited energies that solves
\begin{align}
Q =1 + \frac{\lfloor (\omega-\omega_{\rm gap})\rfloor }{\bar{\omega}} \,,
\end{align}
for a given number $Q$ of electron/hole pairs \cite{Essig:2015cda}. Here, $\lfloor x \rfloor$ is the floor function, which rounds the real number $x$ down to the nearest integer. For silicon, the band gap is given by $\omega_{\rm gap} = 1.11$~eV, and the amount of energy that DM has to deposit in the material to generate an additional electron/hole pair is given by $\bar{\omega}=3.6$ eV \cite{Essig:2015cda}. 

We can now simulate the outcome of a silicon-based DM direct detection experiment operating with an exposure $\xi$ by random sampling the number of observed events with $Q$ electron/hole pairs from a Poisson distribution of mean ${\rm d}\mathscr{R}(Q)/{\rm d} Q |_{\sigma_e=\sigma_e^*; \,\, m_{\chi}=m_\chi^*}$,
where $\sigma_e^*$ and $m_\chi^*$ are the values of the reference cross section and DM particle mass, respectively, that correspond to the observed DM relic density. Here and in all simulations below, we assume $\alpha_{\rm D}=0.5$ and $m_{A'}=3 m_\chi$, which is a standard benchmark point in the LDMX literature. For a given DM particle mass $m_\chi^*$, we read the corresponding value of $\sigma_e^*$ from Fig.~5 of~\cite{LDMX:2018cma}. Finally, for the local DM density we assume $\rho_\chi=0.4$~GeV/cm$^3$, and for $f_\chi(\vec v)$ we assume a truncated Maxwell-Boltzmann distribution with galactic escape speed $v_{\rm esc} = 500$~km~s$^{-1}$, local standard of rest speed $v_0 = 220$~km~s$^{-1}$ and Earth's speed in a reference frame where the mean DM particle velocity is zero, $v_e = 240$~km~s$^{-1}$, which are the default values in {\sffamily DarkELF}~\cite{Knapen:2021bwg}.

Having generated synthetic data for a given exposure, kinetic mixing parameter, and DM particle mass, we use this data to perform Bayesian inference in the ($m_\chi$, $\sigma_e$) plane by employing the multimodal nested sampling algorithm, {\sffamily MultiNest}~\cite{Feroz:2008xx}. To this end, we run {\sffamily MultiNest} with log-uniform prior pdfs for $m_\chi \in [1~\textrm{MeV}, 100~\textrm{MeV}]$ and $\sigma_e \in [10^{-41}~\textrm{cm}^2, 10^{-35}~\textrm{cm}^2]$, setting the number of active points and efficiency parameter to 400 and 1.3, respectively. The outcome of this computation is a two-dimensional posterior pdf in the ($m_\chi$, $\sigma_e$) plane, and one-dimensional marginal posterior pdfs for $m_\chi$ and $\sigma_e$ separately. From these pdfs, we finally calculate means and credible regions for the kinetic mixing parameter and the DM particle mass.

In the Bayesian inference process, we adopt a Poisson likelihood for the direct detection data,
\begin{align}
\mathcal{L}(\{N^i_{\rm exp}\}|m_\chi,\sigma_e) = \prod_{i=1}^{n_Q}\mathcal{L}^{(i)}(N^i_{\rm exp}|m_\chi,\sigma_e) \,,
\end{align}
with
\begin{align}
\mathcal{L}^{(i)}(N^i_{\rm exp}|m_\chi,\sigma_e)= \frac{e^{-N_i(m_\chi,\sigma_e)}}{N^i_{\rm exp}!} N_i(m_\chi,\sigma_e)^{N^i_{\rm exp}}\,,
\end{align}
where $i$ runs from 1 electron/hole pair to $n_Q=10$ electron/hole pairs, $N^i_{\rm exp}$ is sampled from a Poisson distribution of mean ${\rm d}\mathscr{R}(Q)/{\rm d} Q |_{\sigma_e=\sigma_e^*; \,\, m_{\chi}=m_\chi^*;\,\, Q=i}$ as explained above, and, finally, $N_i(m_\chi,\sigma_e)$ is the expected number of events with $Q=i$ that we calculate using the {\sffamily DarkELF} code.

Means and posterior pdfs for $m_\chi$ and $\epsilon$ obtained in this part of the analysis, represent the information one would be able to extract from a future silicon-based DM direct detection experiment that operates with an exposure $\xi$ to test the DM nature of a signal recorded at LDMX.

\subsection{LDMX: direct detection predictions}
\label{sec:ldmx2}

Our analysis in Sec.~\ref{sec:dd} informs us about the region in the ($m_\chi$, $\sigma_e$) plane that would be  ``preferred'' by a silicon-based DM direct detection experiment operating with exposure $\xi$, when the true DM model is characterised by the parameter values $m_\chi^*$ and $\sigma_e^*$. We now want to use this information to predict the expected electron recoil energy and transverse momentum distributions at LDMX based on the recorded (here simulated) direct detection data. 

Let us denote by $f_{\rm DD}(m_\chi,\epsilon)$ the posterior pdf that we obtained from a Bayesian analysis of the direct detection data (see Sec.~\ref{sec:dd}). Using the information encoded in $f_{\rm DD}(m_\chi,\epsilon)$, we can now make the following prediction for the electron recoil energy and transverse momentum distributions that the LDMX experiment should observe, that is $N^{E_e}_{\textrm{DD},i}$ and $N^{P_T}_{\textrm{DD},i}$, $i=1,\dots,n_b$:
\begin{align}
N^{E_e}_{\textrm{DD},i} &= \int {\rm d}m_\chi \int {\rm d} \epsilon\, f_{\rm DD}(m_\chi,\epsilon)  \, N^{E_e}_{i,\textrm{MG}}(m_\chi,\epsilon) \simeq N^{E_e}_{i,\textrm{MG}}(\bar{m}_\chi,\bar{\epsilon}) \nonumber\\
N^{P_T}_{\textrm{DD},i} &= \int {\rm d}m_\chi \int {\rm d} \epsilon\, f_{\rm DD}(m_\chi,\epsilon)  \, N^{P_T}_{i,\textrm{MG}}(m_\chi,\epsilon) \simeq N^{P_T}_{i,\textrm{MG}}(\bar{m}_\chi,\bar{\epsilon})\,, 
\label{eq:meanDD}
\end{align}
where
\begin{align}
\bar{m}_\chi&= \int {\rm d}m_\chi \int {\rm d} \epsilon\, m_\chi \, f_{\rm DD}(m_\chi,\epsilon)\nonumber\\
\bar{\epsilon}&= \int {\rm d}m_\chi \int {\rm d} \epsilon\, \epsilon \, f_{\rm DD}(m_\chi,\epsilon) \,.
\label{eq:null}
\end{align}
Using the information contained in $f_{\rm DD}(m_\chi,\epsilon)$, we can also estimate the associated errors, namely
\begin{align}
\sigma^{E_e}_{\textrm{DD},i} &= \sqrt{\int {\rm d}m_\chi \int {\rm d} \epsilon\, f_{\rm DD}(m_\chi,\epsilon)  \, \left[ N^{E_e}_{\textrm{DD},i} - N^{E_e}_{i,\textrm{MG}}(m_\chi,\epsilon) \right]^2 + \sigma^{E_e\,2}_{\rm MG}}\,, \nonumber\\
\sigma^{P_T}_{\textrm{DD},i} &= \sqrt{\int {\rm d}m_\chi \int {\rm d} \epsilon\, f_{\rm DD}(m_\chi,\epsilon)  \, \left[ N^{P_T}_{\textrm{DD},i} - N^{P_T}_{i,\textrm{MG}}(m_\chi,\epsilon) \right]^2 + \sigma^{P_T\,2}_{\rm MG}}\,,
\label{eq:sigma}
\end{align}
where we have summed in quadrature the error associated with the posterior pdf (first term in the square roots) with the error due to the final size of our Monte Carlo samples, (see Sec.\ref{sec:ldmx}).

\section{Statistical framework}
\label{sec:stat}

Using the strategy and simulation framework introduced in the previous sections, we are now ready to test a hypothetical LDMX signal described by the distributions $N^{E_e}_{\textrm{LDMX},i}$ and $N^{P_T}_{\textrm{LDMX},i}$ against the predicted electron recoil energy and transverse momentum distributions, $N^{E_e}_{\textrm{DD},i}$ and $N^{P_T}_{\textrm{DD},i}$, that we extract from simulated, next generation DM direct detection data. Specifically, we are interested in finding the threshold exposure, $\xi_{\rm th}$, above which one would be able to infer the compatibility, i.e. the common DM origin of the distributions $N^{E_e}_{\textrm{LDMX},i}$ and $N^{E_e}_{\textrm{DD},i}$, as well as $N^{P_T}_{\textrm{LDMX},i}$ and $N^{P_T}_{\textrm{DD},i}$, when they are indeed statistically dependent. Equivalently, the threshold exposure $\xi_{\rm th}$ could be interpreted as follows: for $\xi \le \xi_{\rm th}$ the exposure is so small that the above distributions would appear to be statistically independent, and one would thus erroneously reject the common DM origin of the observed LDMX signal and direct detection data, even in a scenario where the two $E_e$ and $P_T$ distributions are indeed statistically dependent. We perform this test within a statistical framework based on the test statistic\footnote{While the electron recoil energy and transverse momentum distributions are in principle correlated, for simplicity we neglect such correlation here.},
\begin{align}
\chi^2_{\rm TS}= \sum_{i=1}^{n_b} \frac{\left(N^{E_e}_{\textrm{LDMX},i} - N^{E_e}_{\textrm{DD},i}  \right)^2}{N^{E_e}_{\textrm{LDMX},i} + N^{E_e}_{\textrm{DD},i}}
+  \sum_{i=1}^{n_b} \frac{\left(N^{P_T}_{\textrm{LDMX},i} - N^{P_T}_{\textrm{DD},i}  \right)^2}{N^{P_T}_{\textrm{LDMX},i} + N^{P_T}_{\textrm{DD},i}}\,,
\label{eq:chi2}
\end{align}
and the null hypothesis according which LDMX signal and DM direct detection data have a common DM origin. 
When the LDMX data, $N^{E_e}_{\textrm{LDMX},i}$ and $N^{P_T}_{\textrm{LDMX},i}$, and the direct detection data, $N^{E_e}_{\textrm{DD},i}$ and $N^{P_T}_{\textrm{DD},i}$, are sampled from this null hypothesis, i.e.~from Poisson distributions of means 
\begin{align}
\langle N \rangle^{E_e}_{\textrm{LDMX},i}&=\langle N \rangle^{E_e}_{\textrm{DD},i}=N^{E_e}_{i,\textrm{MG}}(m_\chi^*,\epsilon^*)
\, \nonumber\\ 
\langle N \rangle^{P_T}_{\textrm{LDMX},i}& =\langle N \rangle^{P_T}_{\textrm{DD},i}=N^{P_T}_{i,\textrm{MG}}(m_\chi^*,\epsilon^*) 
\,,
\label{eq:means}
\end{align}
$\chi^2_{\rm TS}$'s pdf is a chi-square distribution with $2(n_b-1)$ degrees of freedom in the large sample limit. When using Eq.~(\ref{eq:chi2}) in statistical tests that account for information from the $E_e$ ($P_T$) distribution only, we set to zero the second (first) sum in Eq.~(\ref{eq:chi2}), and assume $n_b-1$ degrees of freedom. 
Within this statistical framework, we define the threshold exposure $\xi_{\rm th}$ as the solution to 
\begin{align}
\chi^2(\xi) = \chi^2_* \,,
\end{align}
where the {\it observed} value of $\chi^2_{\rm TS}$, denoted here and in the figures by $\chi^2$, is given by $\chi^2_{\rm TS}$ evaluated at
\begin{align}
N^{E_e}_{\textrm{LDMX},i} &= N^{E_e}_{i,\textrm{MG}}(m_\chi^*,\epsilon^*)\nonumber\\
N^{P_T}_{\textrm{LDMX},i} &= N^{P_T}_{i,\textrm{MG}}(m_\chi^*,\epsilon^*)\nonumber\\
N^{E_e}_{\textrm{DD},i} &= N^{E_e}_{i,\textrm{MG}}(\bar{m}_\chi,\bar{\epsilon}) \nonumber\\
N^{P_T}_{\textrm{DD},i} &= N^{P_T}_{i,\textrm{MG}}(\bar{m}_\chi,\bar{\epsilon})\,,
\label{eq:chi2obs}
\end{align}
while $\chi^{2}_{*}$ is defined by,
\begin{align}
\int_{\chi^{2}_{*}}^{\infty} {\rm d}x \, f_{n}(x) = 0.05 \,.
\end{align}
Here, $f_n$ is a chi-square pdf with $n$ degrees of freedom, while $n=2(n_b-1)$ or $n=n_b-1$ depending on whether information on both electron recoil energy and transverse momentum is accounted for or not. For $n=2(n_b-1)$, $\chi^{2}_{*}=76.78$, while for $n=n_b-1$ we obtain $\chi^{2}_{*}=42.56$.

In the right hand side of Eq.~(\ref{eq:chi2obs}), we obtain the LDMX observed data from $N^{E_e}_{i,\textrm{MG}}$ and $N^{P_T}_{i,\textrm{MG}}$ evaluated at the true DM particle mass, $m_\chi^*$, and kinetic mixing parameter, $\epsilon^*$: the LDMX signal reflects the true nature of DM. On the other hand, we obtain the DD predicted histograms from $N^{E_e}_{i,\textrm{MG}}$ and $N^{P_T}_{i,\textrm{MG}}$ evaluated at the mean DM mass, $\bar{m}_\chi$, and kinetic mixing parameter, $\bar{\epsilon}$: the DD histograms are predictions based on direct detection data.

For $\chi^2(\xi) < \chi^2_*$ the exposure $\xi$ is large enough to argue that the histograms constructed from $\bar{m}_\chi$ and $\bar{\epsilon}$ and from $m^*_\chi$ and $\epsilon^*$ are statistically compatible, although not necessarily drawn from the same model. Here, by ``statistically compatible'' we mean that the two sets of histograms {\it could} be statistically dependent, but not necessarily\footnote{In particular, even if we fail to reject the null hypothesis, we still cannot confirm to a given statistical significance that the direct detection and LDMX data are from the same DM origin.}. On the other hand, for $\chi^2(\xi) \ge \chi^2_*$ the exposure $\xi$ is so small that our chi-square test would fail to identify a common DM origin of the LDMX and direct detection data (by rejecting the null hypothesis at the 95\% C.L. or higher), even when the two datasets have indeed a common DM origin.  

\begin{figure}[t]
\begin{center}
\includegraphics[width=0.49\textwidth]{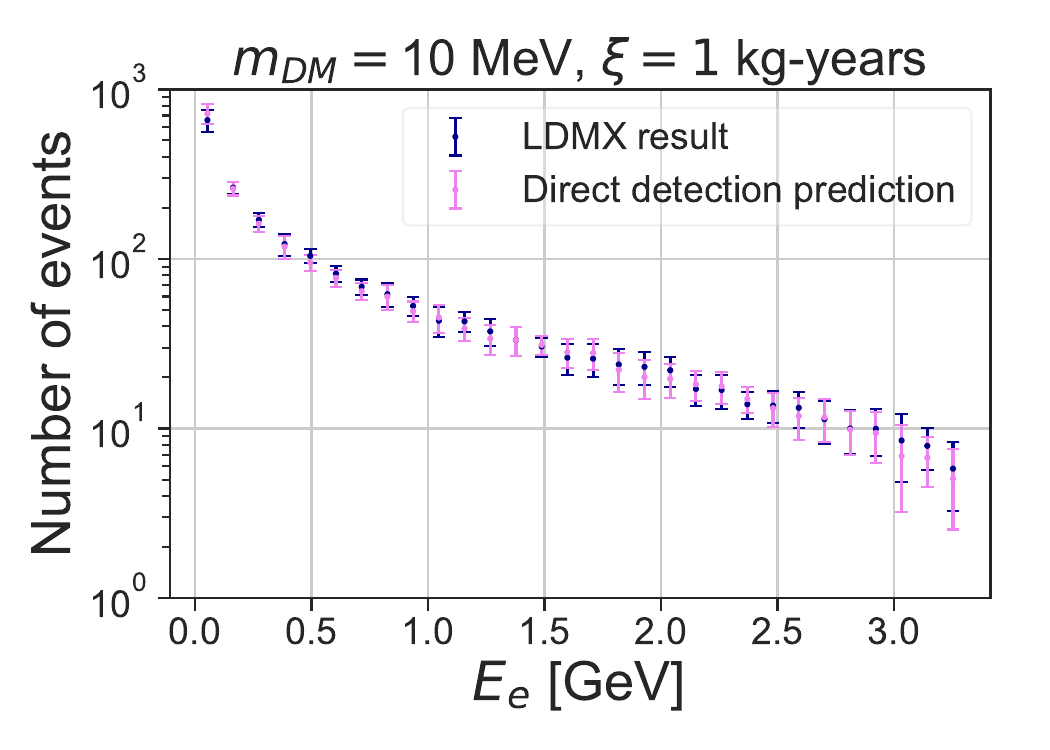}
\includegraphics[width=0.49\textwidth]{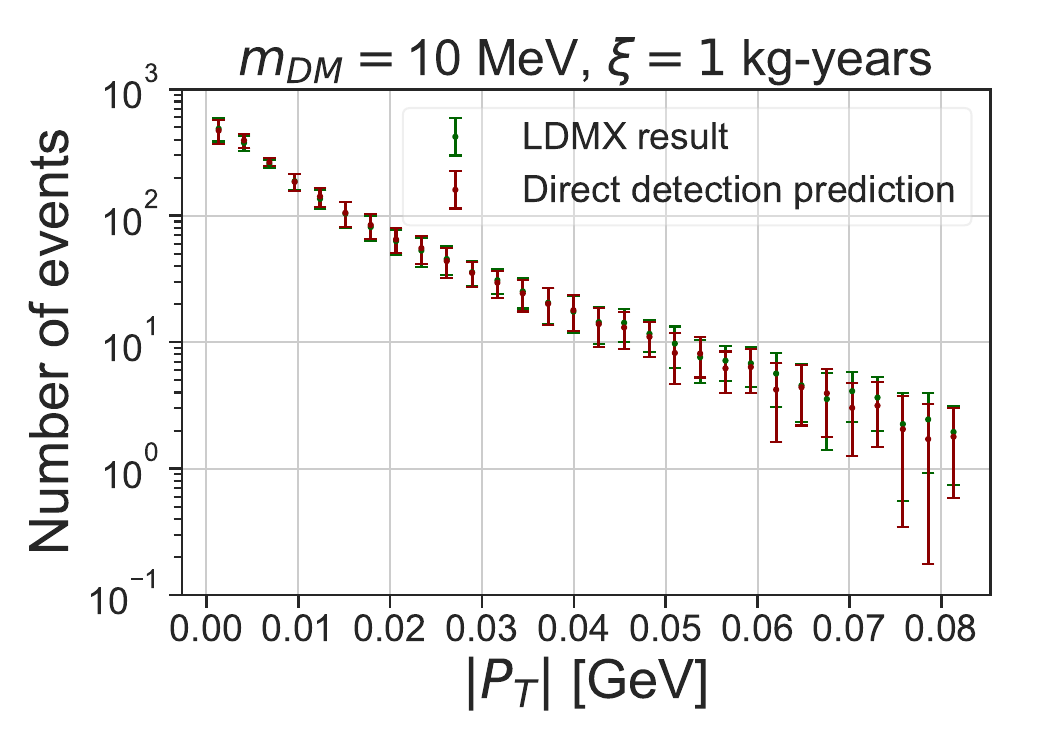}
\includegraphics[width=0.49\textwidth]{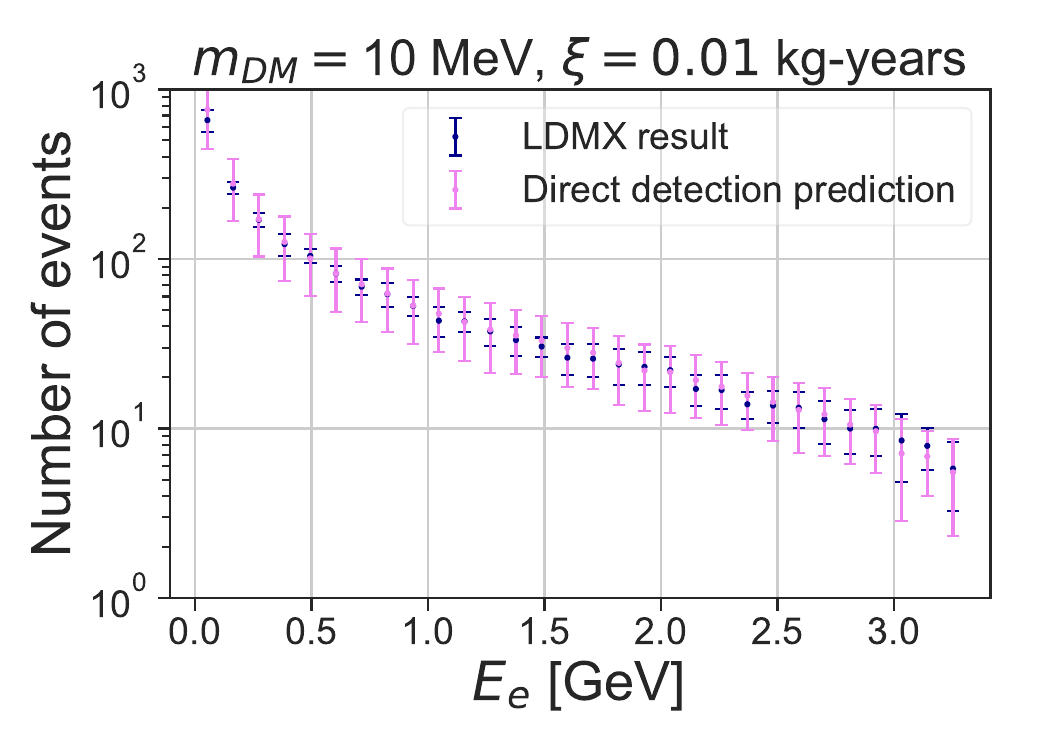}
\includegraphics[width=0.49\textwidth]{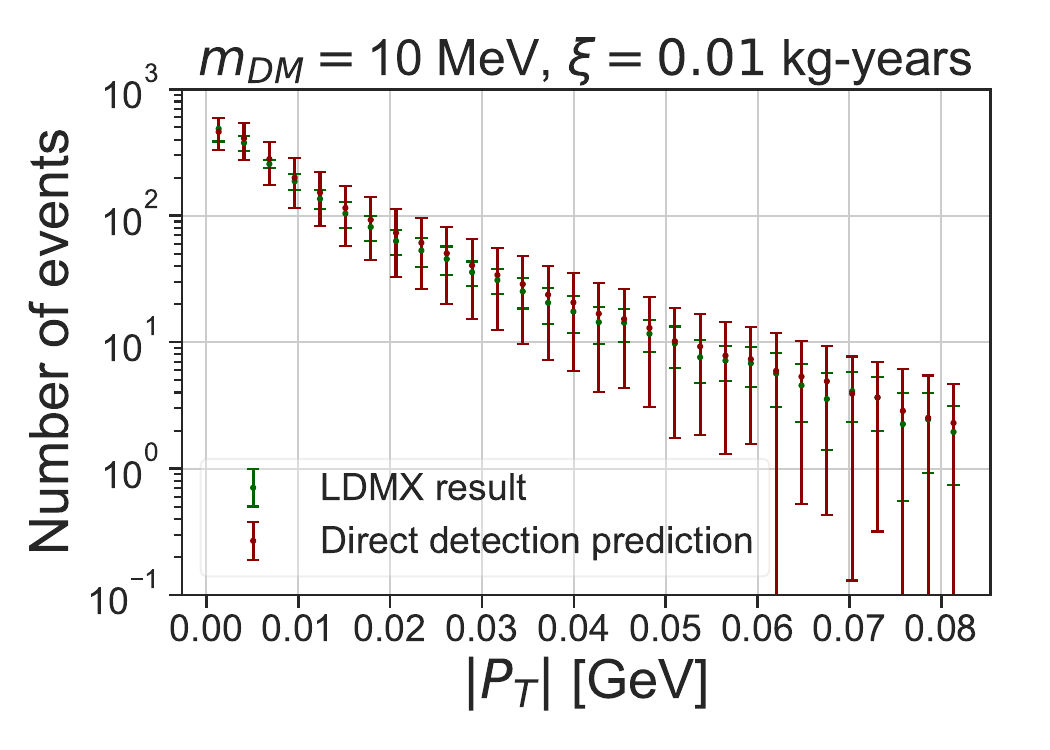}
\caption{{\bf Histograms labelled by ``LDMX result''}: Simulated electron recoil energy and transverse momentum distributions for the benchmark point $m^*_\chi=10$~MeV and $\epsilon^*=4.7\times 10^{-5}$. Simulations are performed with {\sffamily MadGraph} assuming $N_{\rm MG}=10^4$ events per simulation. Finite sample errors are estimated by performing 10 independent simulations, and then calculating mean number of counts and associated standard deviation for each recoil energy and transverse momentum bin. Left (right) panels refer to the simulated recoil energy (transverse momentum) distributions. {\bf Histograms labelled by ``Direct detection prediction''}: Energy and transverse momentum distributions that we predict from the (simulated) data of a future, silicon-based DM direct detection experiment operating with the exposures of $\xi=1$ kg-year (top panels) and $\xi=0.01$ kg-year (bottom panels).
\label{fig:ldmx_counts}}
\end{center}
\end{figure}

\section{Inferring the dark matter nature of an LDMX signal}
\label{sec:results}

We are now ready to estimate the threshold exposure a DM direct detection experiment has to operate with in order to assert the compatibility of a future LDMX signal with its DM origin.

We start by simulating an LDMX signal from a benchmark point in the ($m_\chi$, $\epsilon$) plane. Fig.~\ref{fig:ldmx_counts} shows the simulated electron recoil energy and transverse momentum distributions for the benchmark point $m^*_\chi=10$~MeV and $\epsilon^*=4.7\times 10^{-5}$. The generated events are distributed in 30 energy and momentum bins. In order to estimate the error induced by the finite sample of our Monte Carlo simulations, we perform 10 independent simulations for a given set of DM and {\sffamily MadGraph} parameters, and then calculate mean number of counts and associated standard deviation for each electron recoil energy and transverse momentum bin. The results of this computation correspond to the data points and error bars labelled as ``LDMX result'' in Fig.~\ref{fig:ldmx_counts}, where the left panels refer to the simulated electron recoil energy distributions, whereas the right panels correspond to the simulated transverse momentum data.

\begin{figure}[t]
\begin{center}
\includegraphics[width=0.49\textwidth]{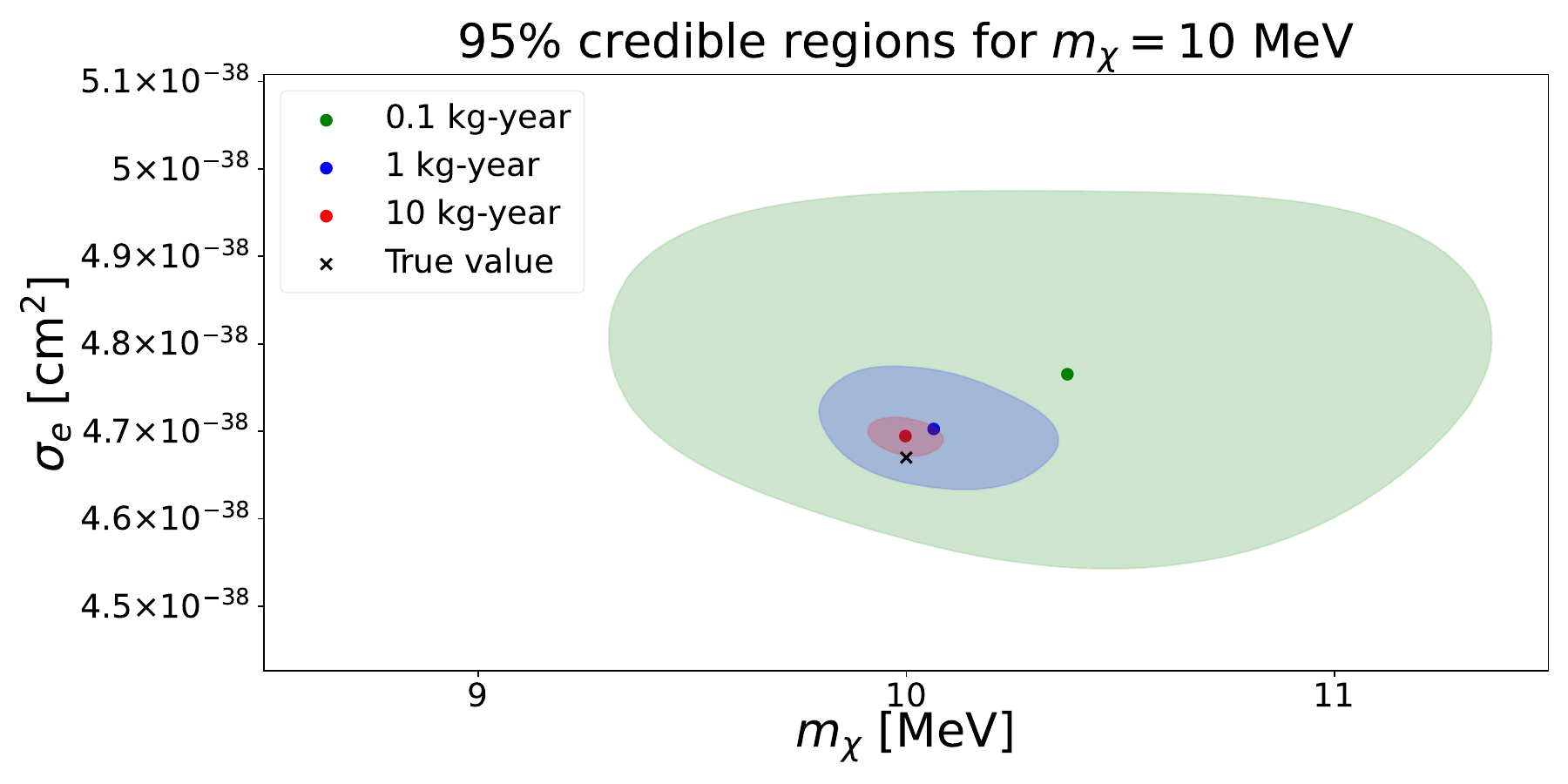}
\includegraphics[width=0.49\textwidth]{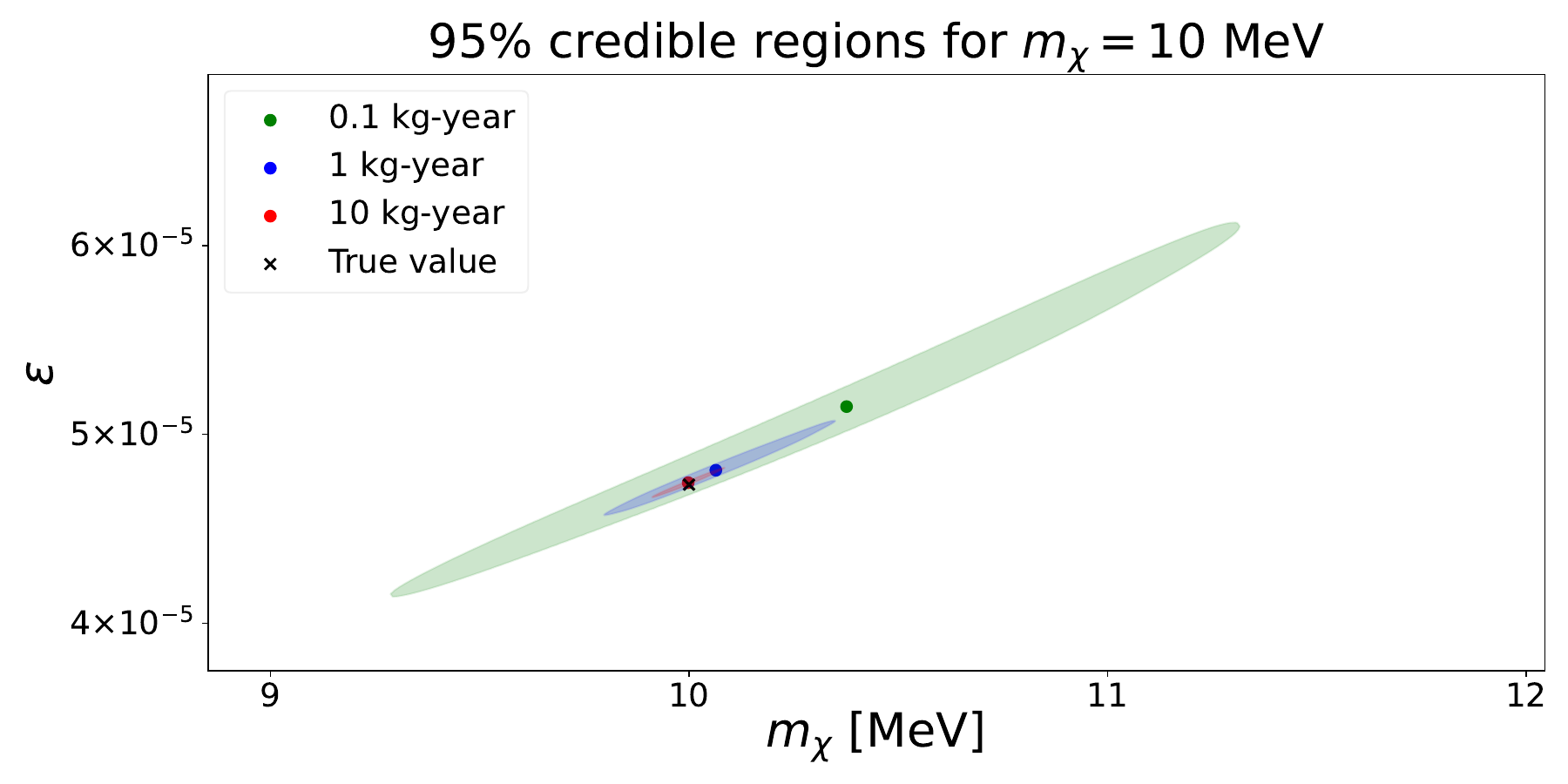}
\caption{Posterior probability density function (pdf) in the ($m_\chi, \sigma_e$) (left) and in the ($m_\chi, \epsilon$) (right)  planes. The pdf is obtained using {\sffamily MultiNest} by fitting the {\sffamily DarkELF} electron/hole pair production rate to simulated data of a next generation DM direct detection experiment. Data are sampled from the benchmark point $m^*_\chi=10$~MeV and $\epsilon^*=4.7\times 10^{-5}$ (corresponding to $\sigma_e=4.67\times 10^{-38}$~cm$^2$), and assume three experimental exposures, namely, $\xi=0.1$~kg-year, $\xi=1$~kg-year and  $\xi=10$~kg-year. Benchmark point and predicted posterior means are superimposed to the associated 95\% credible regions. 
\label{fig:pdf}}
\end{center}
\end{figure}

In Fig.~\ref{fig:ldmx_counts}, we also compare the LDMX signal we simulate as described above and in Sec.~\ref{sec:ldmx} with the energy and transverse momentum distributions that we predict from the (simulated) data of a future, silicon-based DM direct detection experiment operating with the exposures of 1 kg-year (top panels) and 0.01 kg-year (bottom panels). To obtain this prediction, first we run {\sffamily DarkELF} to calculate the expected rate of electron-hole pair production in a silicon target for $m^*_\chi=10$~MeV and $\epsilon^*=4.7\times 10^{-5}$ (corresponding to $\sigma_e=4.67\times 10^{-38}$~cm$^2$). This is the same benchmark point from which the LDMX signal has been simulated. Second, from the electron-hole pair production rate expected at this benchmark point we random sample the number of events with $Q=1,\dots,10$ electron-hole pairs a direct detection experiment with $\xi=0.01$~kg-year and $\xi=1$~kg-year would observe.~We then fit the {\sffamily DarkELF} pair production rate to the simulated data by using {\sffamily MultiNest} to obtain a posterior pdf in the ($m_\chi$, $\epsilon$) plane, as explained in Sec.~\ref{sec:dd}. As an example of our {\sffamily MultiNest} runs, Fig.~\ref{fig:pdf} shows the posterior pdf that we find for three values of $\xi$. Finally, we use this pdf to calculate the mean predicted DM mass and mixing parameter, and Eqs.~(\ref{eq:Ee}) and (\ref{eq:sigma})
to obtain the $E_e$ and $P_T$ histograms and associated errors shown in Fig.~\ref{fig:ldmx_counts}. From this figure, we can already conclude that the recoil energy and transverse momentum distributions observed at LDMX and the analogous ones extracted from the direct detection data are qualitatively comparable for $\xi=1$~kg-year, although a quantitative comparison requires the statistical methods of Sec. \ref{sec:stat}, as we discuss below.

\begin{figure}[t]
\begin{center}
\includegraphics[width=0.49\textwidth]{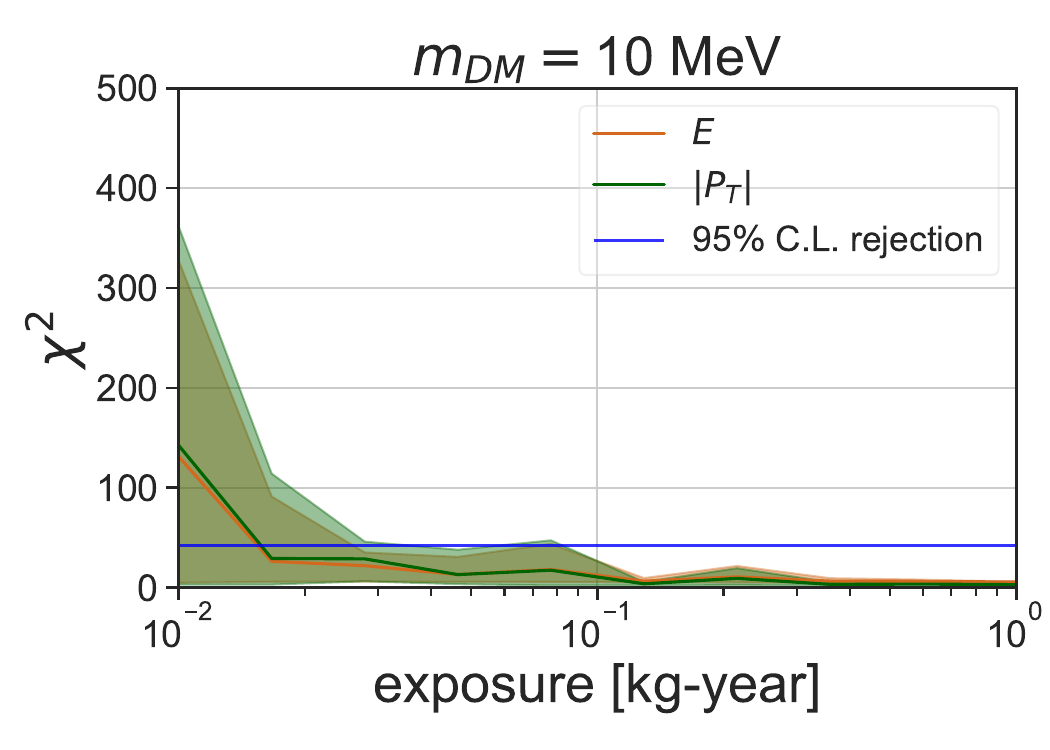}
\includegraphics[width=0.49\textwidth]{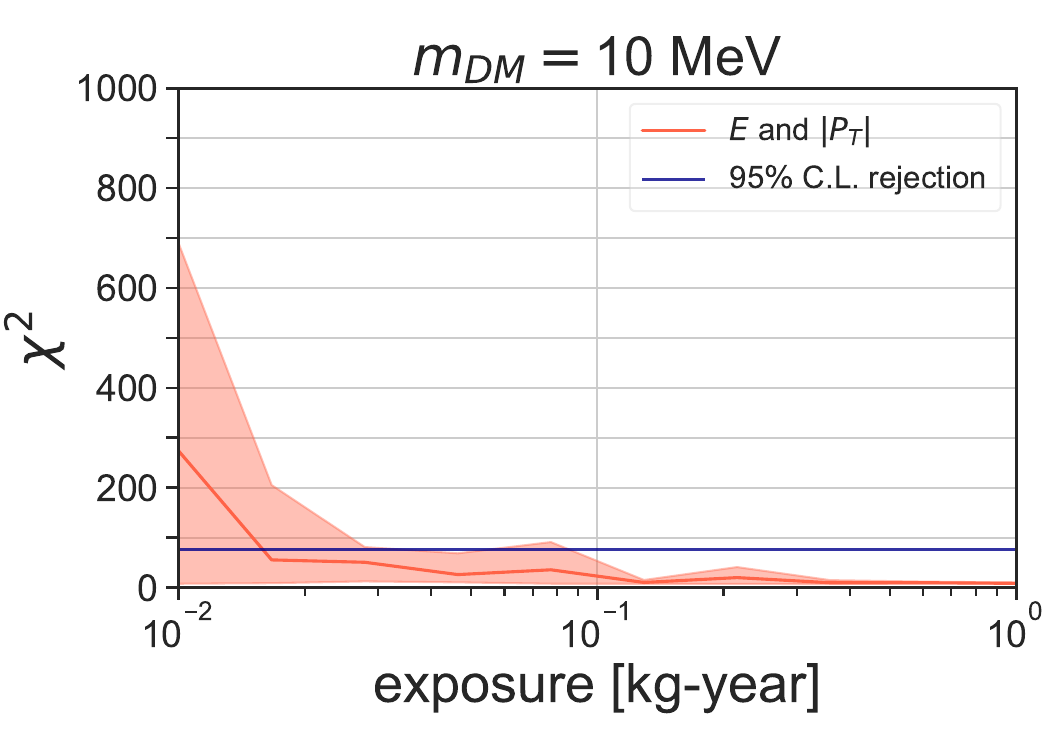}
\caption{Comparison between the ``observed'' recoil energy and transverse momentum distributions $N^{E_e}_{\textrm{LDMX},i}$ and  $N^{P_T}_{\textrm{LDMX},i}$, and the predicted distributions $N^{E_e}_{\textrm{DD},i}$ and $N^{P_T}_{\textrm{DD},i}$. The comparison is made by computing the chi-square test statistic, $\chi^2$, in Eq.~(\ref{eq:chi2}) as a function of the experimental exposure $\xi$ for different analysis settings. {\bf Left panel}: Results obtained by considering the information contained in the recoil energy and transverse momentum data separately. {\bf Right panel}: $\chi^2$ as a function of $\xi$ assuming that information on both the recoil energy and transverse momentum distributions is taken into account. For all settings, we calculate $\chi^2$ at a given $\xi$ for 10 different realisations of the DM direct detection data, and then compute the mean value of $\chi^2$ and associated standard deviation over these 10 realisations. In both panels of Fig.~\ref{fig:EpT}, the horizontal blue line corresponds to $\chi^{2}_*$, defined as in Sec.~\ref{sec:stat}. For $\chi^2>\chi^2_*$, one would reject the common DM origin of LDMX and DM direct detection data at 95\% C.L. For $\chi^2<\chi^2_*$, the direct detection exposure is large enough to rightfully identify the common DM origin of DM direct detection and LDMX data, when the two datasets are statistically dependent.
\label{fig:EpT}}
\end{center}
\end{figure}

\begin{figure}[t]
\begin{center}
\includegraphics[width=0.49\textwidth]{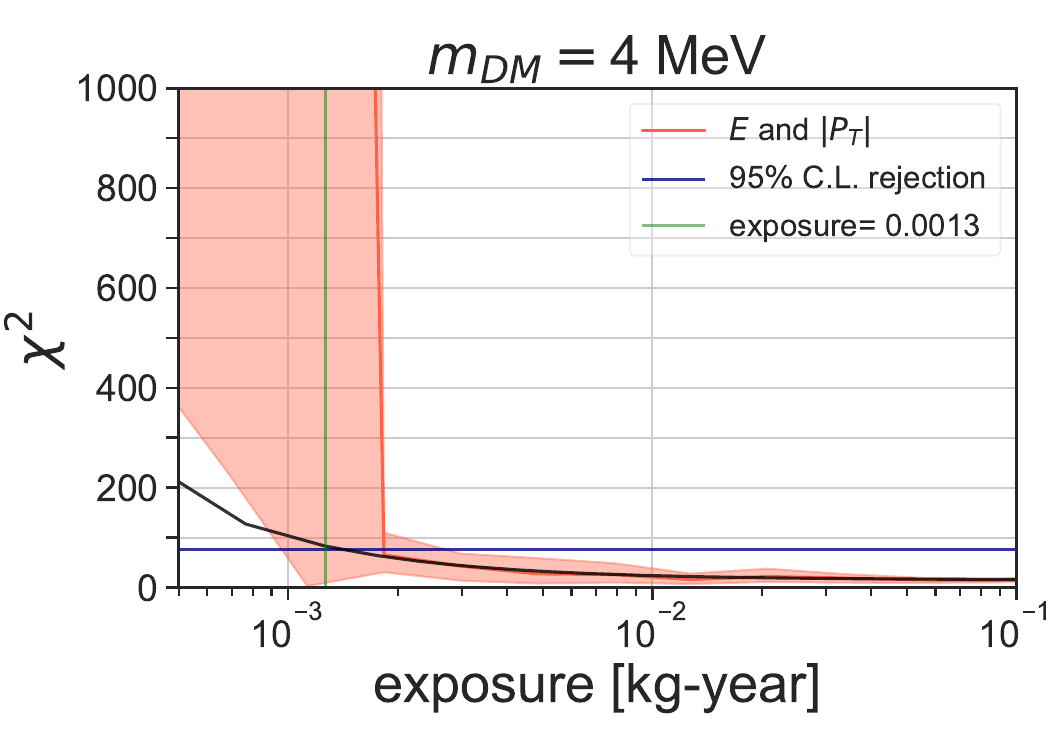}
\includegraphics[width=0.49\textwidth]{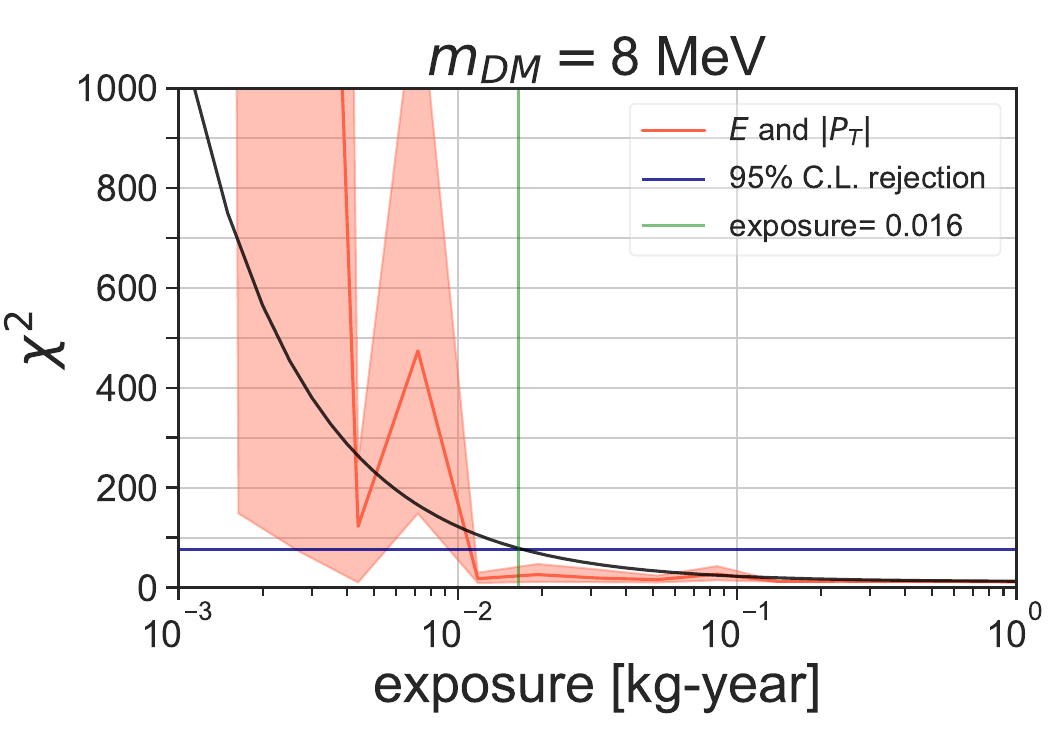}
\includegraphics[width=0.49\textwidth]{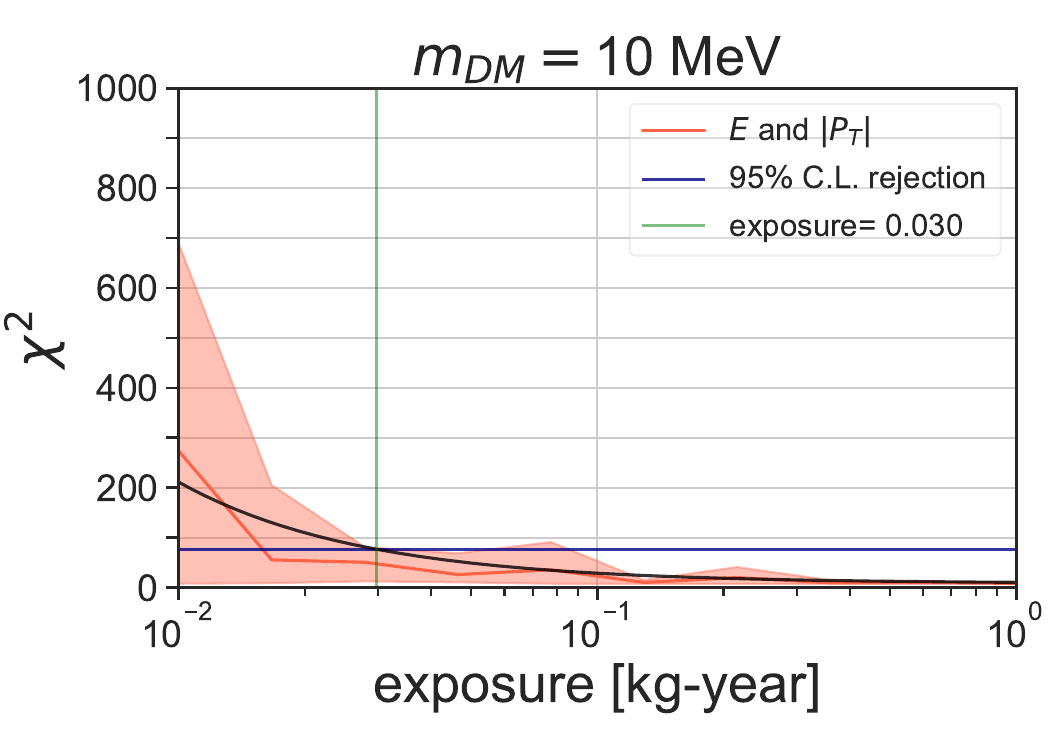}
\includegraphics[width=0.49\textwidth]{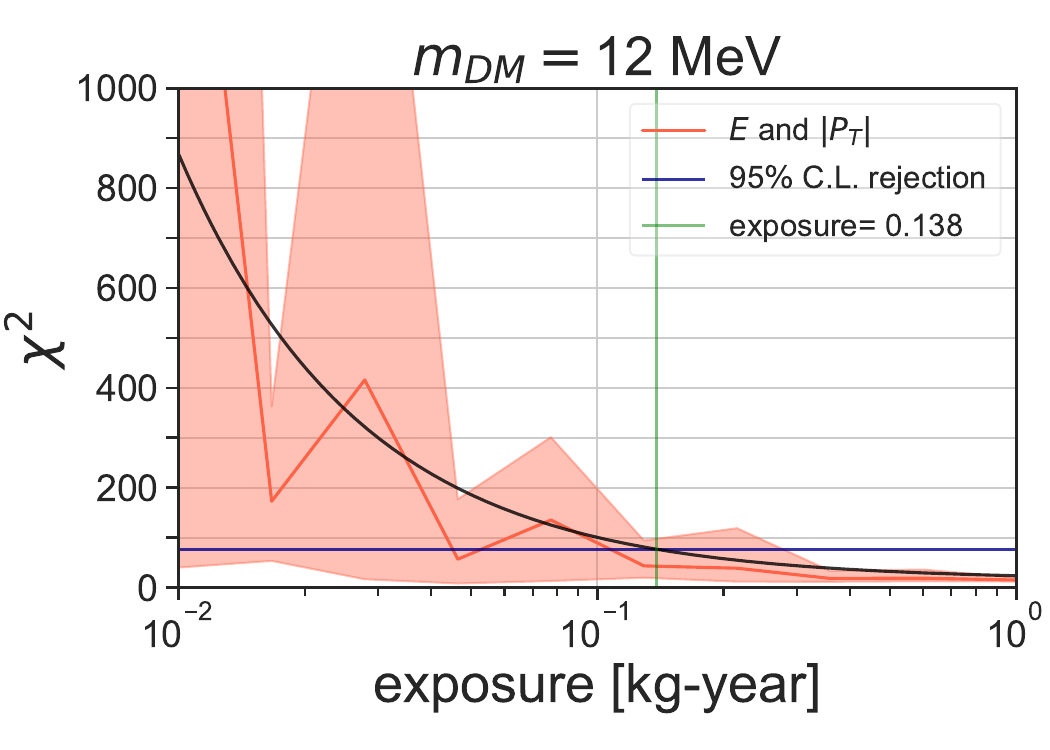}
\includegraphics[width=0.49\textwidth]{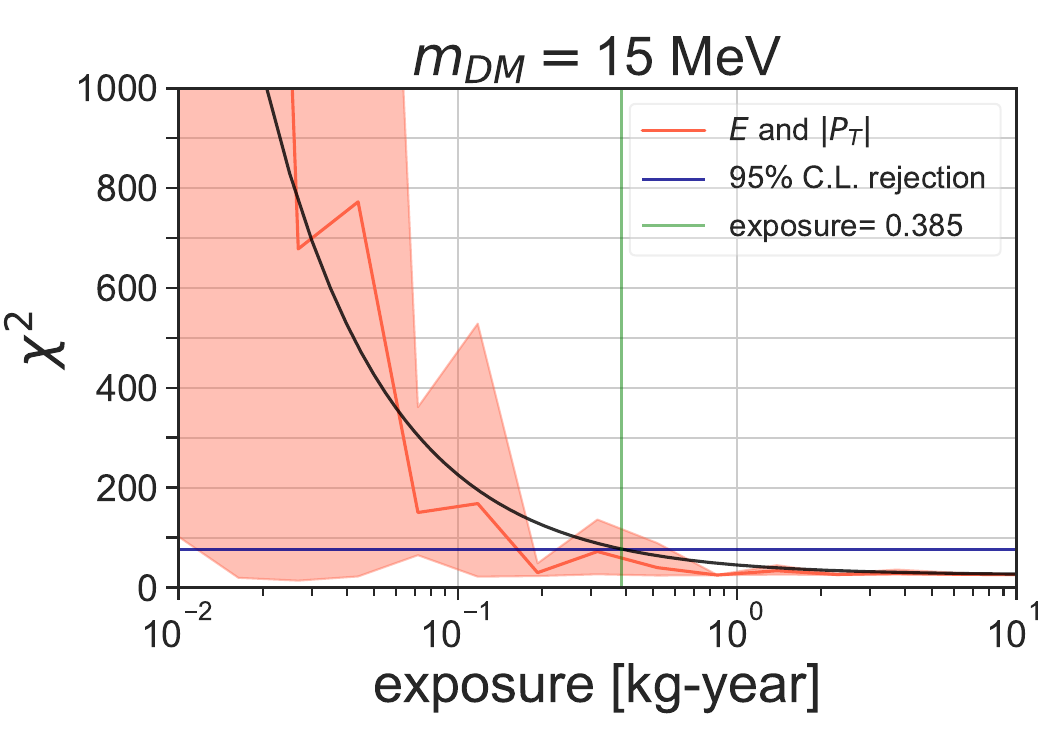}
\includegraphics[width=0.49\textwidth]{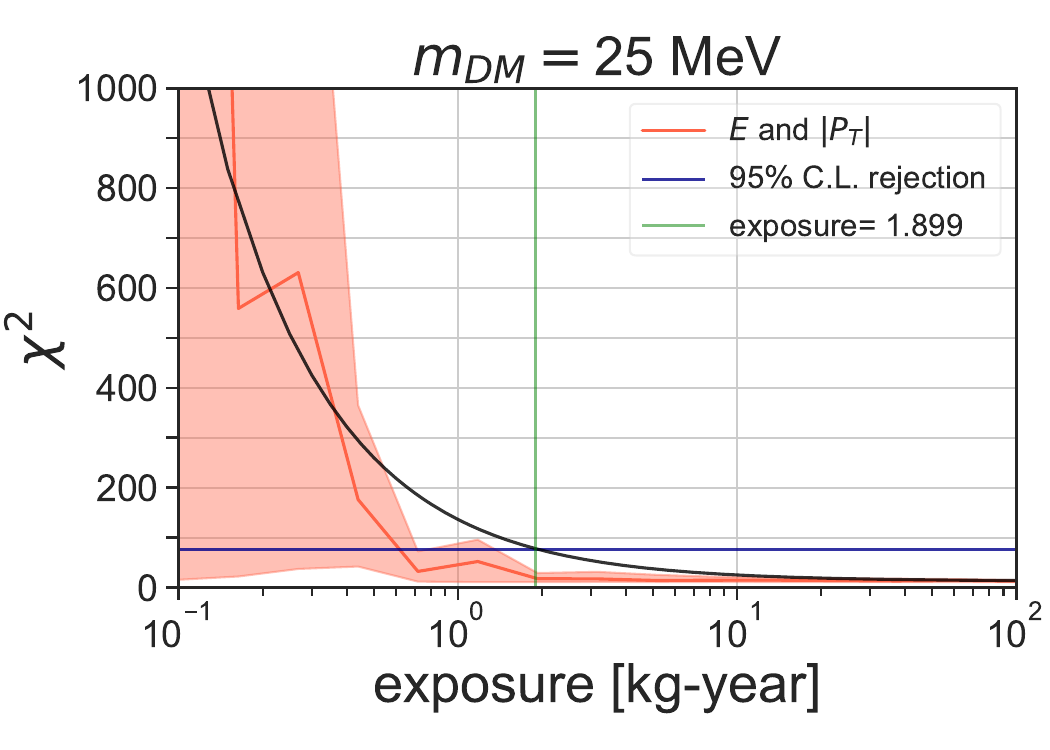}
\caption{Chi-square test statistic, $\chi^2$ in Eq.~(\ref{eq:chi2}), as a function of the experimental exposure $\xi$ for six values of the DM particle mass, ranging from 4~MeV to 25~MeV. In all panels, a vertical green line denotes our estimated threshold experimental exposure, $\xi_{\rm th}$, which we obtain from the intersection of the black line associated with a fit to the mean $\chi^2(\xi)$ curve and the blue horizontal line, corresponding to $\chi^2_*$. Above $\xi_{\rm th}$, one would be able to infer the compatibility, i.e. the common DM origin of the distributions $N^{E_e}_{\textrm{LDMX},i}$ and $N^{E_e}_{\textrm{DD},i}$, as well as $N^{P_T}_{\textrm{LDMX},i}$ and $N^{P_T}_{\textrm{DD},i}$, when they are indeed statistically dependent. Equivalently, for $\xi \le \xi_{\rm th}$ the exposure is so small that the above distributions would appear to be statistically independent, even when the two $E_e$ and $P_T$ distributions are indeed statistically dependent.  
The value we obtain for $\xi_{\rm th}$ grows with the DM particle mass, and varies from 0.0013~kg-year for a DM particle mass of $m_\chi=4$~MeV up to 1.9~kg-year for a DM mass $m_\chi=25$ MeV.
\label{fig:all}}
\end{center}
\end{figure}

\begin{figure}[t]
\begin{center}
\includegraphics[width=0.49\textwidth]{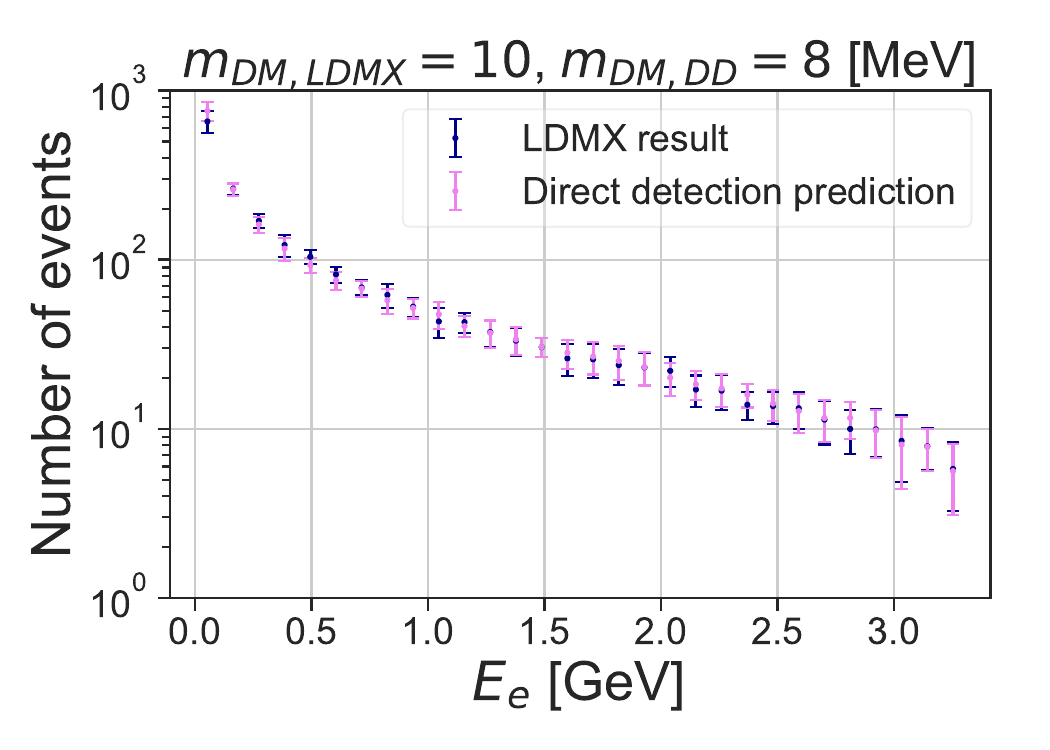}
\includegraphics[width=0.49\textwidth]{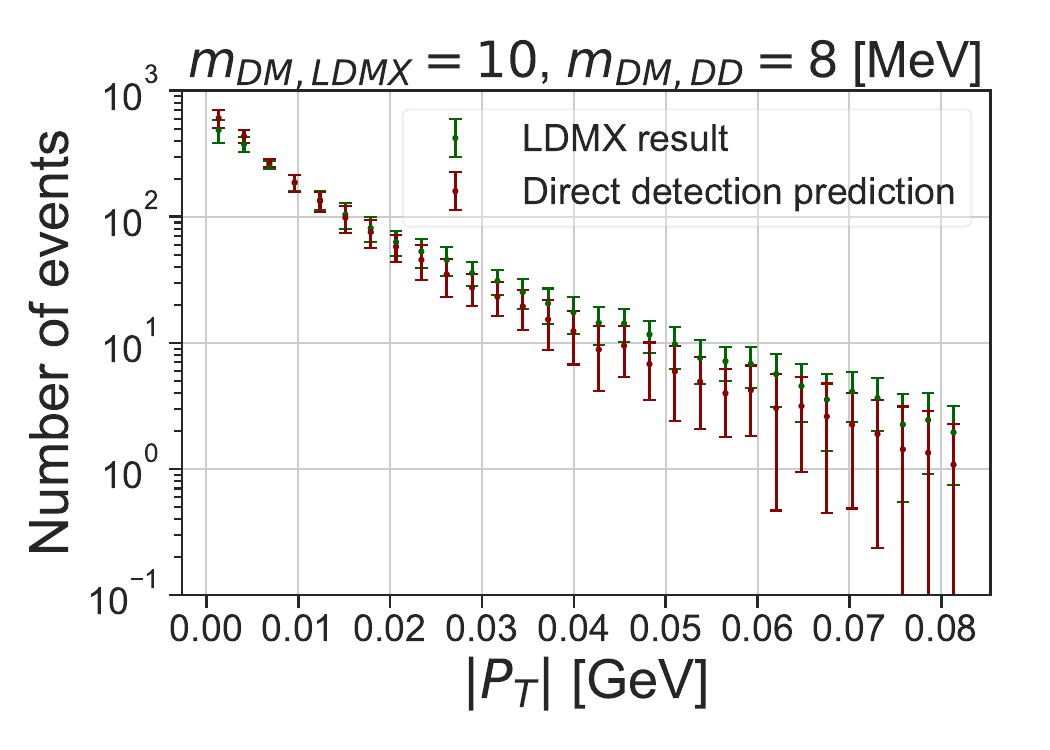}
\includegraphics[width=0.49\textwidth]{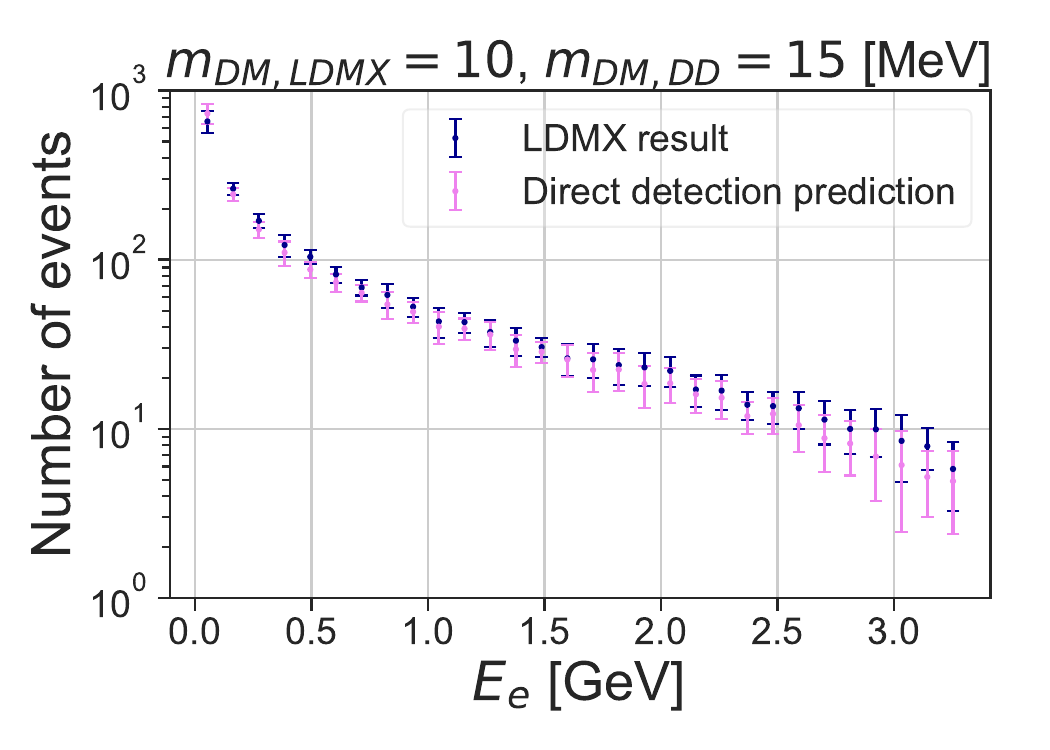}
\includegraphics[width=0.49\textwidth]{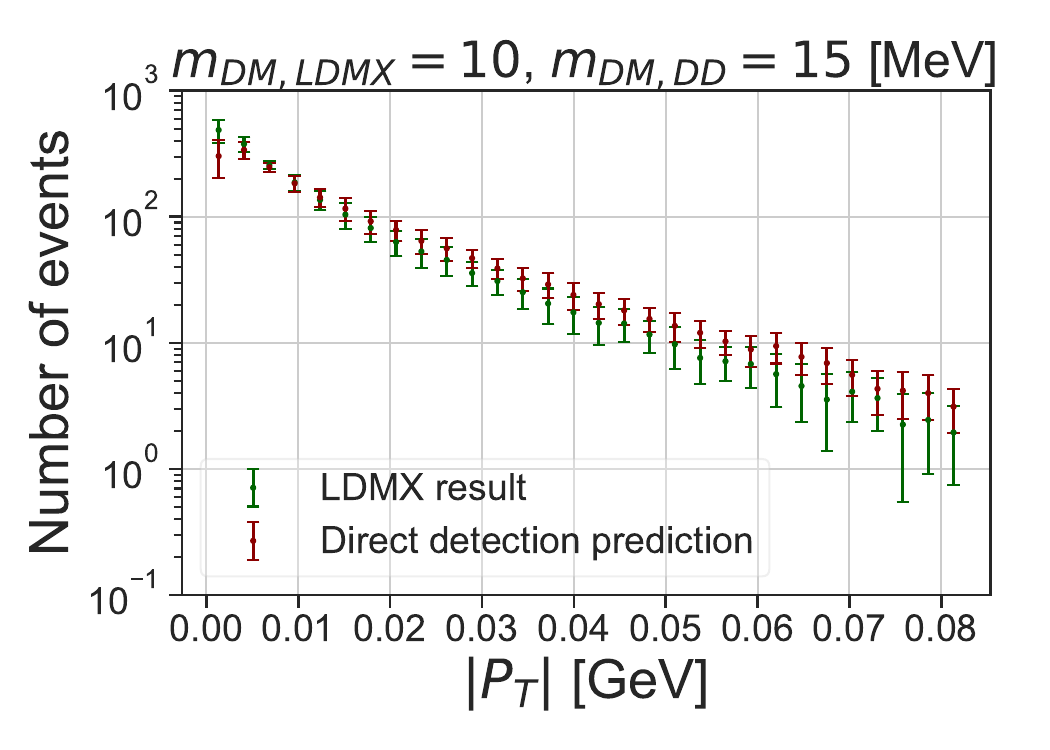}
\caption{Same as Fig.~\ref{fig:ldmx_counts}, now assuming different masses for the LDMX, $m_{\rm DM, LDMX}$, and direct detection, $m_{\rm DM, DD}$, simulations.
\label{fig:ldmx_counts2}}
\end{center}
\end{figure}

\begin{figure}[t]
\begin{center}
\includegraphics[width=0.49\textwidth]{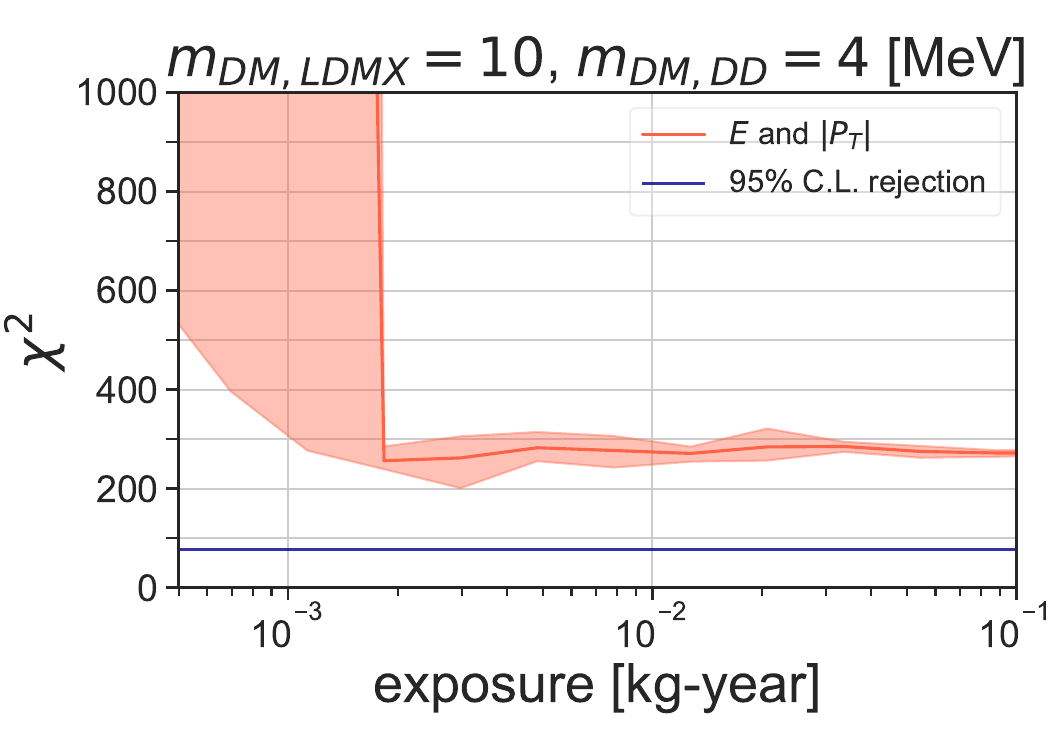}
\includegraphics[width=0.49\textwidth]{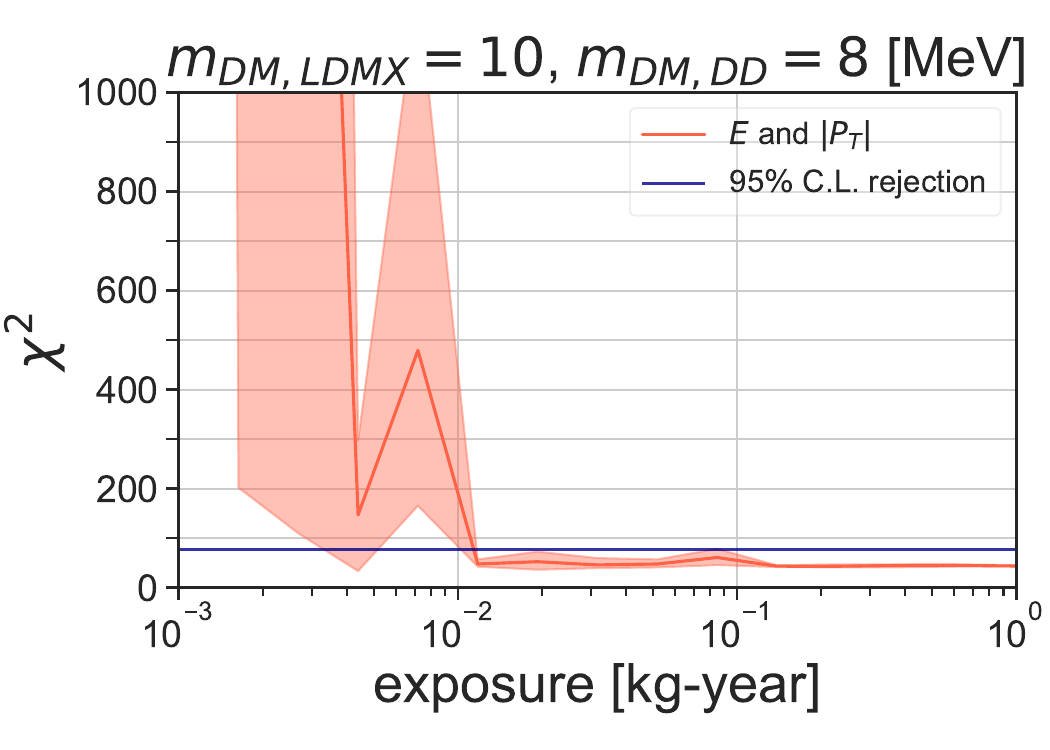}
\includegraphics[width=0.49\textwidth]{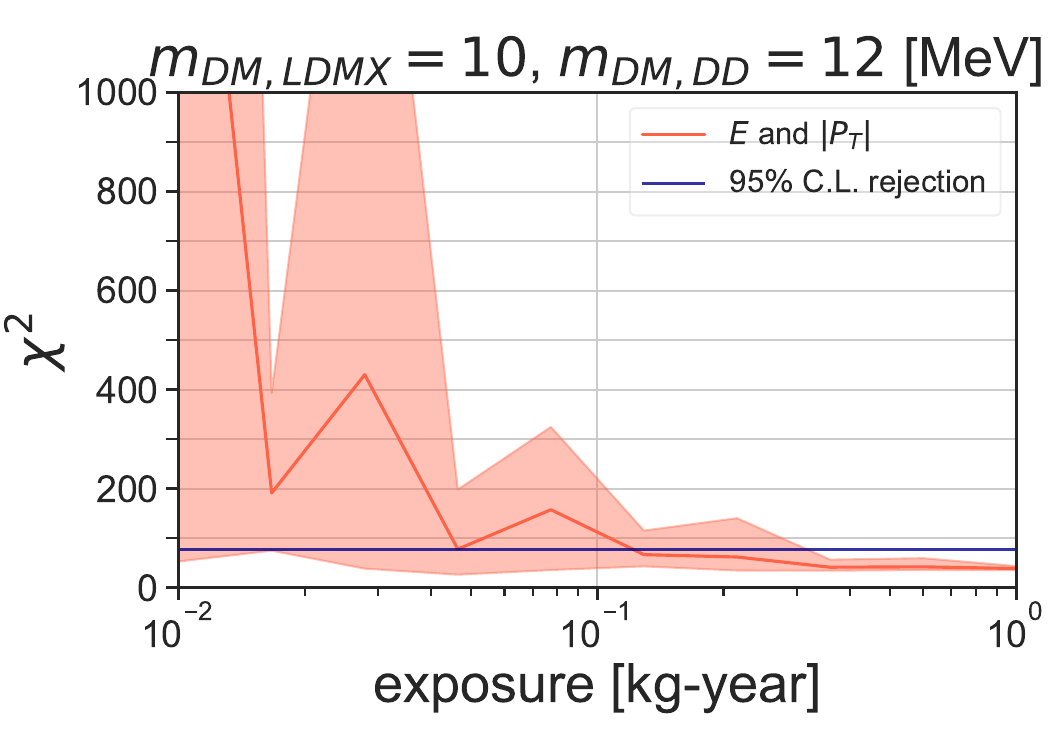}
\includegraphics[width=0.49\textwidth]{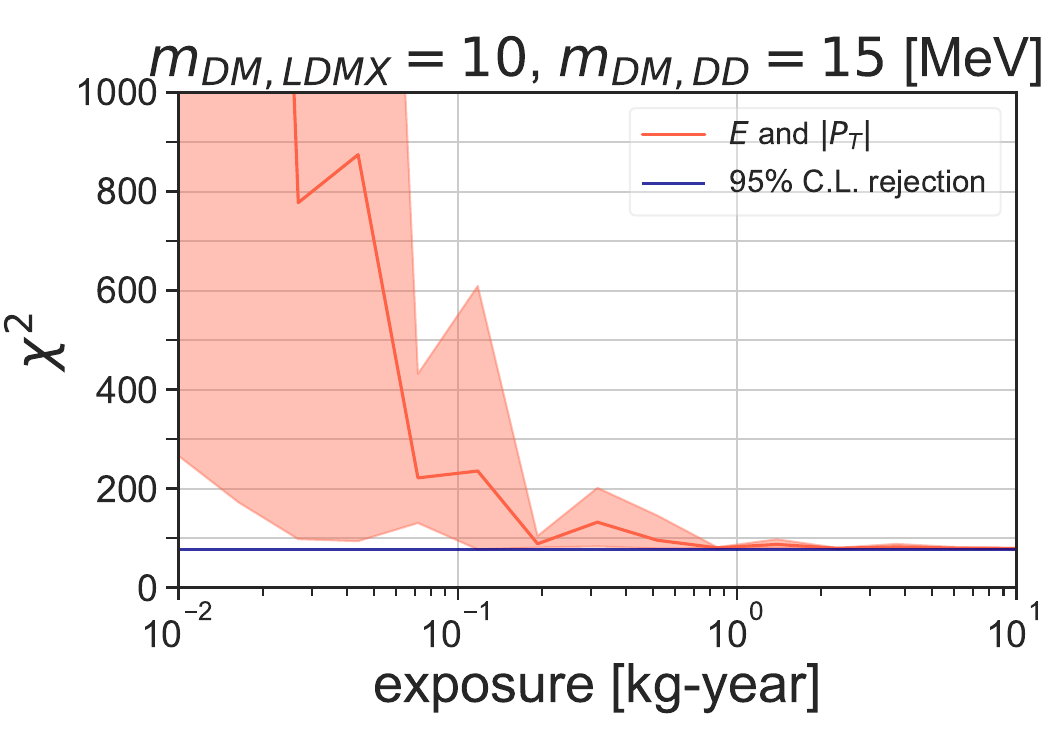}
\caption{Same as Fig.~\ref{fig:all}, now assuming different masses for the LDMX, $m_{\rm DM, LDMX}$, and direct detection, $m_{\rm DM, DD}$, simulations. As expected, $\chi^2$ is always above the blue line (corresponding to $\chi_*^2$) with the exception of the top right panel, where the DM mass used in the LDMX and direct detection simulations are fairly close.
\label{fig:all2}}
\end{center}
\end{figure}

\begin{figure}[t]
\begin{center}
\centering
\includegraphics[width=0.49\linewidth]{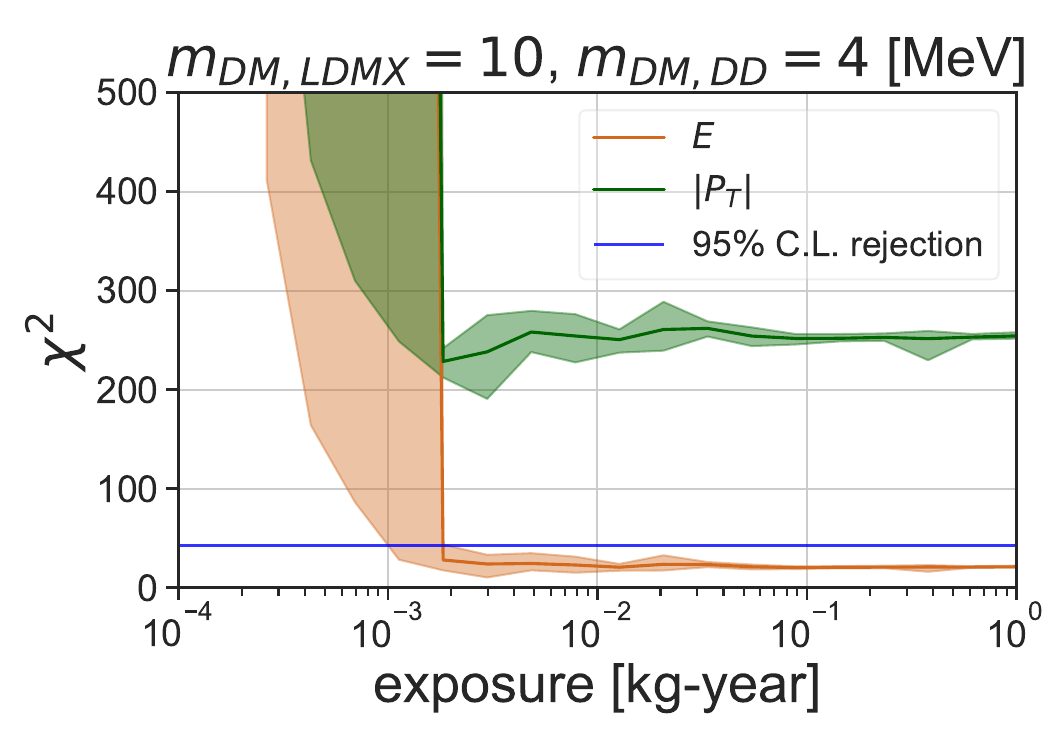}
\includegraphics[width=0.49\linewidth]{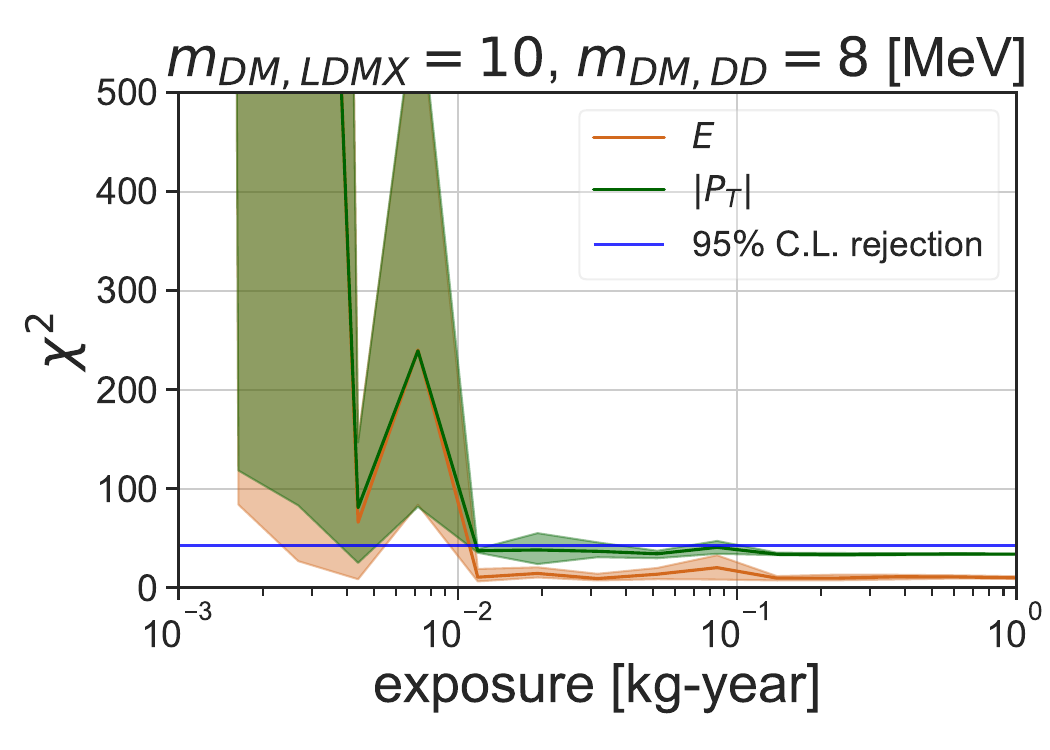}
\includegraphics[width=0.49\linewidth]{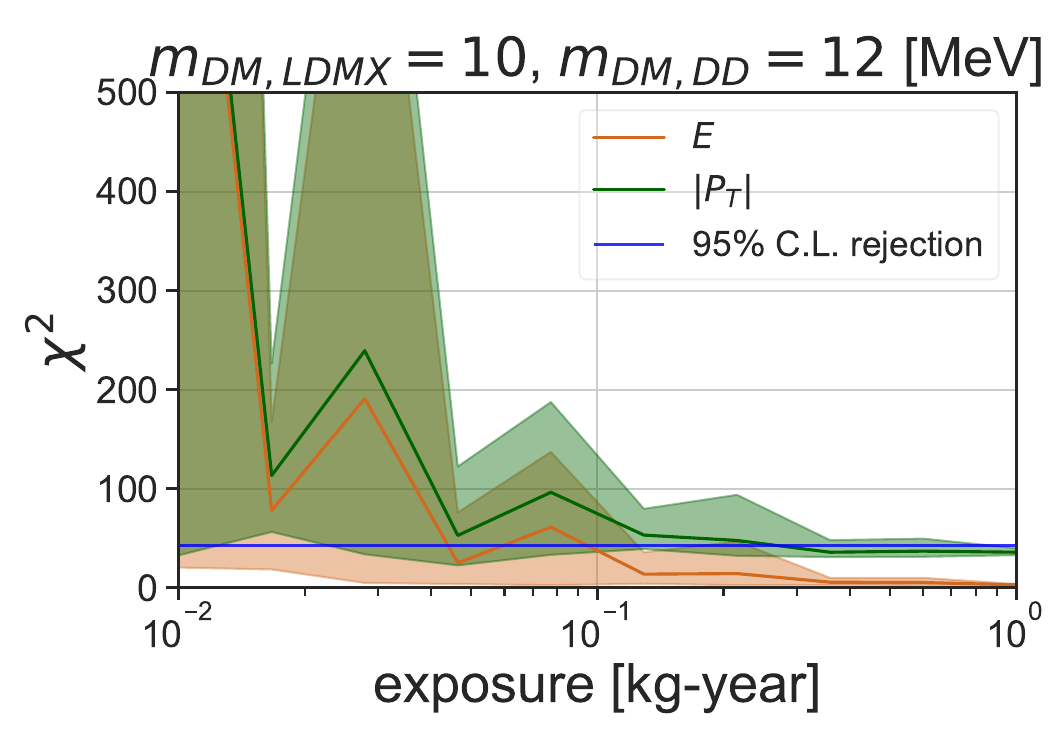}
\includegraphics[width=0.49\linewidth]{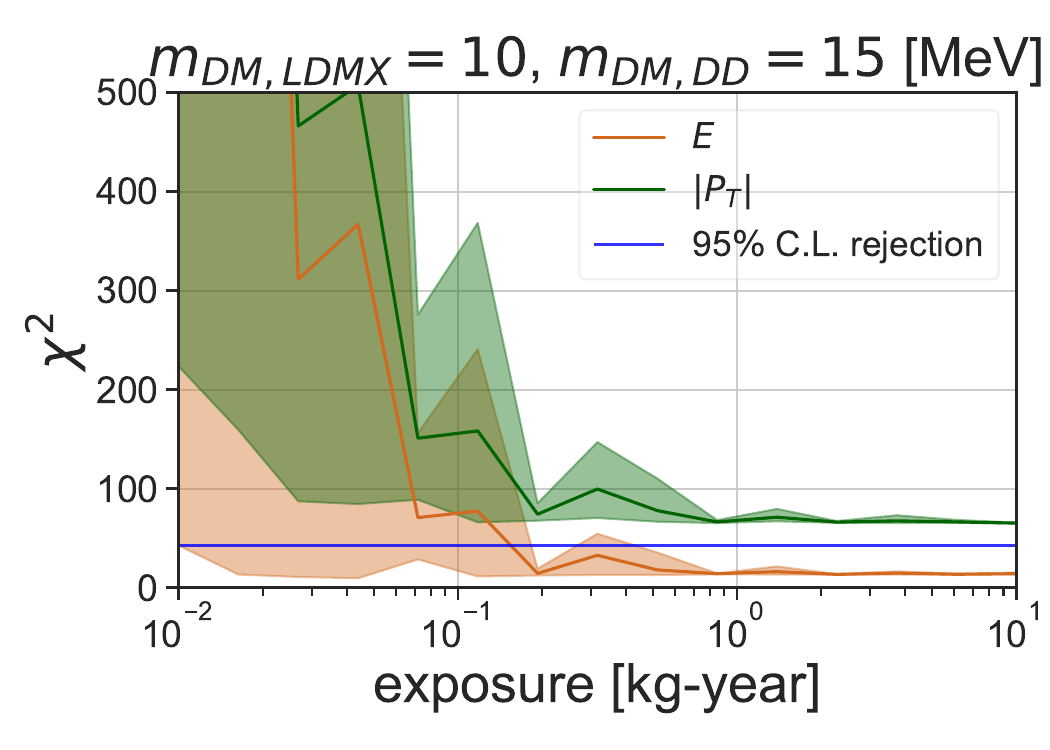}
\caption{Same as Fig. \ref{fig:all2} but with the contributions from the energy and $p_T$ to the $\chi^2$ plotted separately, with the contribution from $E$ in orange and from $p_T$ in green.}
\label{fig:chi2_10vs15EpT}
\end{center}
\end{figure}

We now compare the ``observed'' energy and transverse momentum distributions $N^{E_e}_{\textrm{LDMX},i}$ and  $N^{P_T}_{\textrm{LDMX},i}$ with the predicted counterparts $N^{E_e}_{\textrm{DD},i}$ and $N^{P_T}_{\textrm{DD},i}$ by using the test statistic in Eq.~(\ref{eq:chi2}). The results of this comparison are reported in Fig.~\ref{fig:EpT}, which shows $\chi^2$ as a function of the experimental exposure $\xi$ for three different analysis settings. In the left panel, we report results obtained by considering the information contained in the recoil energy and transverse momentum data separately (i.e. first and second setting, respectively). The right panel shows the $\xi$-dependence of $\chi^2$ when information on both the recoil energy and transverse momentum distributions is taken into account (third analysis setting). As anticipated in Sec.~\ref{sec:stat}, for the first two settings the effective number of degrees of freedom is $(n_b-1)$, for the third one is $2(n_b-1)$. For all settings, we calculate $\chi^2$ at a given $\xi$ for 10 different realisations of the DM direct detection data, and then compute the mean value of $\chi^2$ and associated standard deviation over these 10 realisations. This allows us to report in Fig.~\ref{fig:EpT} mean $\chi^2(\xi)$ and an associated error band for each experimental setting (the colour code and the distribution type is indicated in the legends). In both panels of Fig.~\ref{fig:EpT}, the horizontal blue line corresponds to $\chi^{2}_*$, defined as in Sec.~\ref{sec:stat}.

For $\xi < \xi_{\rm th}\sim 0.02$~kg-year, $\chi^2>\chi^2_*$, and the distributions $N^{E_e}_{\textrm{LDMX},i}$ and $N^{P_T}_{\textrm{LDMX},i}$ appear to be statistically independent from the distributions $N^{E_e}_{\textrm{DD},i}$ and $N^{P_T}_{\textrm{DD},i}$. We would in this case reject the common DM origin of the LDMX signal and direct detection data, even in a scenario where these distributions are indeed statistically dependent. We therefore need to operate a DM direct detection experiment with an experimental exposure that is larger than $\xi_{\rm th}\sim0.02$~kg-year to avoid erroneously rejecting the DM origin of a hypothetical LDMX signal, and infer the common DM origin of the distributions, e.g. $N^{E_e}_{\textrm{LDMX},i}$ and $N^{E_e}_{\textrm{DD},i}$, when they are indeed statistically dependent. From Fig.~\ref{fig:EpT} we can also conclude that both the $E_e$ and the $P_T$ distributions play an important role in determining the threshold exposure $\xi_{\rm th}$, with the recoil energy distribution being slightly more informative for the specific benchmark point that has been considered in the figure.

So far, we focused on a benchmark point in the ($m_\chi$, $\epsilon$) parameter space where $m_\chi=10$~MeV. In Fig.~\ref{fig:all}, we explore how the function $\xi \rightarrow \chi^2(\xi)$ changes when the DM particle mass varies from 4~MeV to 25~MeV\footnote{This DM mass range is chosen because below $\sim 4$~MeV, direct detection sensitivity drops significantly \cite{Essig:2011nj}, and above $\sim 25$~MeV we expect the Migdal effect to become relevant \cite{Knapen:2021bwg}.}.
In all panels, a vertical green line denotes our estimated threshold experimental exposure, $\xi_{\rm th}$, which we extract from the intersection of the blue horizontal line, corresponding to $\chi^2_*$, and the black line associated with a fit of functional form $A+B/\xi$\footnote{This scaling law is justified because of the expected scaling of $\text{Var}[N_{DD,i}^{E}]$ with respect to the exposure, which goes as $1/\xi$. In principle $\chi^2$ will approach a larger value than what we compute here, i.e. the fit will give a non-zero $A$ parameter. This is partly due to when evaluating $\chi^2$ we neglect statistical fluctuations in $N^{E_e}_{\textrm{LDMX},i}$ and $N^{P_T}_{\textrm{LDMX},i}$, which are fixed at the value they take for $m_\chi=m_\chi^*$ and $\epsilon=\epsilon^*$, as explained in Sec.~\ref{sec:stat}. Also note that if the underlying models causing the direct detection and LDMX signals are different, such as having different masses, the $\chi^2$ will approach a larger non-zero number -- justifying the use of $A$ in the fitting function.} to the mean $\chi^2(\xi)$ curve. The value we obtain for $\xi_{\rm th}$ grows with the DM particle mass, and varies from 0.0013~kg-year for a DM particle mass of $m_\chi=4$~MeV to 1.9~kg-year for $m_\chi=25$ MeV. A difference of three orders of magnitude in $\xi_{\rm th}$ can be understood by realizing that the sensitivity to $\epsilon$ of DM direct detection experiments grows by a bit over one order of magnitude when varying $m_\chi$ from 25 MeV to 4 MeV, and that this sensitivity scales with the exposure as $1/\sqrt{\xi}$.  
The standard deviation of the $\chi^2$ samples, as shown in Figs.~\ref{fig:all} and \ref{fig:EpT}, are subject to fairly large statistical fluctuations associated with the stochastic output of our {\sffamily MadGraph} simulations. In order to produce smoother plots, we would have to increase significantly the number of {\sffamily MadGraph} simulations per point in parameter space, which unfortunately goes beyond our current computational capabilities.

Imagine now that the distributions $N^{E_e}_{\textrm{LDMX},i}$ and $N^{P_T}_{\textrm{LDMX},i}$ are indeed statistically independent from the distributions $N^{E_e}_{\textrm{DD},i}$ and $N^{P_T}_{\textrm{DD},i}$. We can emulate this scenario, by sampling $N^{E_e}_{\textrm{LDMX},i}$ and $N^{P_T}_{\textrm{LDMX},i}$ from a benchmark point with $m_\chi=10$~MeV while $N^{E_e}_{\textrm{DD},i}$ and $N^{P_T}_{\textrm{DD},i}$ from a second benchmark point corresponding to a different DM particle mass. Would the formalism based on the test statistic in Eq.~(\ref{eq:chi2}) allow us to identify the statistical independence of the two distributions, and thus reject the hypothesis that they have been drawn from the same model at 95\% C.L. or higher? We address this question in Fig.~\ref{fig:ldmx_counts2} and Fig.~\ref{fig:all2}. Fig.~\ref{fig:ldmx_counts2} compares the energy distributions $N^{E_e}_{\textrm{LDMX},i}$ and $N^{E_e}_{\textrm{DD},i}$ (left panels), and the transverse momentum distributions $N^{P_T}_{\textrm{LDMX},i}$ and $N^{P_T}_{\textrm{DD},i}$ (right panels) for $\xi=1$~kg-year, and for $m_\chi=8$~MeV (top panels) and $m_\chi=15$~MeV (bottom panels). One can appreciate significant differences in the two sets of distributions, especially when comparing the transverse momentum distributions obtained by assuming $m_\chi = 10$ MeV at LDMX and $m_\chi = 8$ MeV in the direct detection simulations. These qualitative differences are confirmed quantitatively from the results of a chi-square test reported in Fig.~\ref{fig:all2}. In particular, the formalism based on the test statistics in Eq.~(\ref{eq:chi2}) allows us to rightfully reject the hypothesis that the LDMX and direct detection distributions have been drawn from the same model at 95\% C.L. or higher for all exposures when the DM mass underlying the direct detection data is either 15 MeV or 4 MeV. At the same time, $\chi^2$ crosses $\chi^2_*$ for some finite value of the experimental exposure when the compared DM masses are close, as in the top right and bottom left panels in Fig.~\ref{fig:all2}.

The energy distribution and the $p_T$ distribution do not contribute equally to the $\chi^2$. The difference is shown in Fig. \ref{fig:chi2_10vs15EpT}, where each contribution is plotted separately. It was noted above that for the scenario of equal masses for the direct detection and LDMX simulations, the energy of the recoil electron is slightly more informative. In the case of different masses, as in Fig. \ref{fig:all2}, the $p_T$ contributes more to the $\chi^2$, thus providing the most information regarding whether or not the two experimental signals are statistically independent.

Uncertainties in the parameters that govern the DM space and velocity distribution, such as the local DM density and galactic escape velocity, could be included in our analysis as nuisance parameters. This could be done by: 1) adding a Gaussian factor to our direct detection likelihood for each astrophysical parameter to be accounted for; 2) marginalizing the resulting likelihood over all nuisance parameters. Based on the impact of astrophysical uncertainties on the DM-induced excitation rate in semiconductor detectors found in, e.g., \cite{Hryczuk:2020trm} (see in particular Fig.~\ref{fig:EpT} right panel) we expect that this procedure would broaden the posterior pdfs of our Fig.~\ref{fig:pdf} by an order one factor in the DM-electron scattering cross section direction, without shifting significantly the associated means. Since we obtain our predicted electron recoil energy and transverse momentum distributions from such pdf means, as shown in the right-hand-side of Eq.~(\ref{eq:meanDD}), we do not expect a major impact of astrophysical uncertainties on our conclusions. At the same time, a quantitative assessment would require a detailed analysis which goes beyond the scope of the present work.

\section{Summary and conclusion}
\label{sec:conclusions}
A signal in upcoming fixed target experiments, such as LDMX, would require an independent validation to assert its DM origin. In this work, we have explored the possibility of using next generation direct detection experiments to shed light on the origin of a hypothetical LDMX signal. In particular, we have computed the threshold exposure a silicon-based DM direct detection experiment has to operate with to assert the statistical compatibility of the LDMX signal with the electron/hole pair data reported at the next generation direct detection experiment. 

Specifically, we have proposed a four-step analysis strategy to test the DM origin of an LDMX signal. In the first step, the hypothetical LDMX signal is recorded (in our case, simulated). In the second step a DM direct detection experiment operates with increasing exposure to test the DM origin of the LDMX signal. In the third step, the electron/hole pair data reported by the DM direct detection experiment is analysed, and a posterior pdf for the DM particle mass and mixing parameter is extracted from the data. The posterior pdf obtained from the DM direct detection data is used to predict the expected electron recoil energy and transverse momentum distributions at LDMX. In the last step, indirectly predicted (from direct detection data) and directly observed (at LDMX) electron recoil energy and transverse momentum distributions are compared in a chi-square test to assert whether LDMX signal and direct detection data are statistically compatible. By ``statistically compatible'', we mean that the predicted and observed number of counts per energy and per transverse momentum bin {\it can} be statistically dependent, although not necessarily.

Quantitatively, we find that the threshold experimental exposure grows with the DM mass, and varies from 0.0013~kg-year for a DM mass of $m_\chi=4$~MeV to 1.9~kg-year for $m_\chi=25$ MeV. These levels of exposure are within reach at future DM direct detection experiments \cite{Mitridate:2022tnv}. For comparison, the current exposure of DAMIC is 85.23~g-days (corresponding to 0.00023 kg-year \cite{DAMIC-M:2023gxo}).

Our strategy to test the DM origin of a LDMX signal highlights the strong complementarity of DM searches at direct detection experiments and new mediator particle searches at LDMX.

\acknowledgments 
RC acknowledges support from an individual research grant from the Swedish Research Council (Dnr.~2022-04299). RC and TG have also been funded by the Knut and Alice Wallenberg Foundation, and performed their research within the ``Light Dark Matter'' project (Dnr. KAW 2019.0080).
\bibliographystyle{JHEP}
\bibliography{ref,ref2}

\end{document}